\documentclass[iop,twocolappendix,numberedappendix]{emulateapj}

\usepackage{natbib}
\usepackage[loose, ugly]{units}
\usepackage[english]{babel}
\usepackage{blindtext}
\usepackage{hyperref}

\newcommand*{\del}{\ensuremath{\partial}}

\newcommand{\ceff}{\ensuremath{c_{\rm eff}}}
\newcommand{\hI}{H~\textsc{i}}
\newcommand{\hmol}{\ensuremath{\mathrm{H}_2}}

\shorttitle{Gravitationally Unstable Disk Galaxies}
\shortauthors{N. J. Goldbaum, M. R. Krumholz, and J. C. Forbes}

\addto\extrasenglish{%

}

\begin{document}

%\renewcommand\arraystretch{1.2}


\title{Mass Transport and Turbulence in Gravitationally Unstable Disk
  Galaxies. I: The Case of Pure Self-Gravity}

\author{
  Nathan J. Goldbaum,\altaffilmark{1,2}
  Mark R. Krumholz,\altaffilmark{1}
  John C. Forbes,\altaffilmark{1}
}

\altaffiltext{1}{
  National Center for Supercomputing Applications, University of Illinois, 
  1205 West Clark Street, Urbana-Champaign, Illinois 61820, USA}
\altaffiltext{2}{
  Department of Astronomy \& Astrophysics, University of California,
  Santa Cruz, CA 95064, USA}

\begin{abstract}
  The role of gravitational instability-driven turbulence in determining the
  structure and evolution of disk galaxies, and the extent to which gravity
  rather than feedback can explain galaxy properties, remains an open question.
  To address it, we present high resolution adaptive mesh refinement simulations
  of Milky Way-like isolated disk galaxies, including realistic heating and
  cooling rates and a physically motivated prescription for star formation, but
  no form of star formation feedback. After an initial transient, our galaxies reach
  a state of fully-nonlinear gravitational instability. In this state, gravity drives turbulence and
  radial inflow. Despite the lack of feedback, the gas in our galaxy models
  shows substantial turbulent velocity dispersions, indicating that
  gravitational instability alone may be able to power the velocity dispersions
  observed in nearby disk galaxies on \unit[100]{pc} scales. Moreover, the rate
  of mass transport produced by this turbulence approaches $\sim\, 1$ $M_\odot$
  yr$^{-1}$ for Milky Way-like conditions, sufficient to fully fuel star
  formation in the inner disks of galaxies. In a companion paper we add feedback
  to our models, and use the comparison between the two cases to understand what
  galaxy properties depend sensitively on feedback, and which can be understood
  as the product of gravity alone. All of the code, initial conditions, and
  simulation data for
  our model are publicly available.
\end{abstract}


\section{Introduction}

\subsection{Gravitational Instability as a Driver of Galactic Evolution}
\label{motivation}

Until the past few years, most theoretical work on galaxy evolution has focused
on reproducing observed correlations in the bulk properties of galaxies, such as
the stellar mass-halo mass relation \citep[e.g.,][]{moster13, behroozi13}, the
stellar mass-star formation relation \citep[e.g.,][]{daddi07, elbaz07,
  noeske07}, and the mass-metallicity relation \citep[e.g.,][]{tremonti04,
  pilyugin04}. In such correlations, galaxies are treated as single points, and
theoretical models have been largely content to treat them as such. In these
models, the behavior of galaxies is almost entirely dictated by a balance
between cosmological accretion and feedback \citep[e.g.,][]{bouche10, lilly13,
  forbes14b, mitra15}, and for this reason most theoretical attention has been
focused on stellar feedback as the dominant driver of galaxy evolution.

The recent availability of large samples of spatially-resolved maps of gas,
metals, and star formation in nearby galaxies has opened up a new frontier in
the study of galaxies: explaining the radial distribution of these
quantities. Observational studies conducted to date have turned up puzzling
facts that demand explanation, which cannot obviously be explained simply by a
local balance of accretion versus feedback at all locations in a
galaxy. Instead, these observations point to a potentially significant role for
radial redistribution of material as a key process for understanding the
evolution and growth of galaxies over cosmological timescales.

The first observational puzzle concerns the radial distribution of gas and star
formation in present-day disk galaxies. Observations show that the neutral gas
in such galaxies is distributed with an approximately universal exponential
profile, with typical gas scale lengths of $\sim\, 0.5\, r_{25}$, where $r_{25}$ is
the optical radius of the galaxy \citep{regan01, schruba11, bigiel12}. The gas
is dominated by \hmol\ inside $\sim\, 0.4\, r_{25}$, while \hI\ predominates
at larger radii \citep[e.g.,][]{leroy08}. The radial distribution of star
formation is, to first approximation, simply linearly proportional to the \hmol\
distribution \citep{bigiel08, schruba11, leroy13}, so star formation is much
more radially concentrated than the total neutral gas, and predominantly occurs
inside $\sim\, 0.5\, r_{25}$. In such regions, the time required to convert all the
available gas to stars is $\sim\, 2$ Gyr, much less than a Hubble time
\citep{bigiel08, leroy13}. Thus the inner parts of galaxies will become
gas-depleted and cease star formation unless fresh gas is supplied at an equal
rate, which for $L_*$ galaxies is a few $M_\odot$ yr$^{-1}$
\citep{kennicutt12}. While some disk galaxies, including the Milky Way, do have
inner gas holes and quenched central star formation, such systems appear to be
the exception rather than the rule among the local disk galaxy population
\citep{bigiel12}.

How the fresh gas responsible for the lack of quenched centers can be explained
is an unsolved problem. While condensation of gas from the hot halo
\citep{marinacci10, fraternali13, hobbs13} will supply gas in a centrally
concentrated manner, direct accretion of cold gas from the intergalactic medium
\citep[IGM; e.g.,][]{keres05, dekel09} primarily deposits gas at large
galactocentric radii, far from the actively star-forming regions that need to be
refueled. This can lead to build-up of gas at large radii
\citep[c.f.][]{dutton12}. Gas recycling from evolved stellar populations can
potentially provide some of the required mass \citep{leitner11} at small
galactocentric radii, but, in addition to uncertainties about whether this
channel provides enough mass and is consistent with various chemical evolution
constraints, the radial distribution of recycled gas has not been explored. In
summary, the lack of quenched, gas-depleted centers in local disk galaxies seems
to require some form of gas redistribution within the disk, for which
gravitational instability is an obvious candidate.

A second surprising observation regarding the radial distribution of gas
concerns gas velocity dispersions. \hI\ lines in disk galaxies show
super-thermal velocity dispersions of $\sim\, 10$ \unit{km s$^{-1}$}
\citep[e.g.,][]{vanzee99, petric07, tamburro09, ianjamasimanana12,
  ianjamasimanana15}. Velocity dispersions are highest toward the centers of
galaxies, but they decline only shallowly with radius, and remain superthermal
even well outside $r_{25}$. While star formation feedback alone appears able to
drive the observed velocity dispersion in inner disks, the same is not true
outside $r_{25}$, where the star formation rate drops precipitously but the
velocity dispersion does not \citep{tamburro09, ianjamasimanana15}.  Dwarf
galaxies represent an extreme in this regard: due to their very low rates of
star formation, supernova feedback cannot plausibly provide enough energy to
drive the observed turbulence \textit{anywhere} within them \citep{stilp13}.
Nor do magneto-rotational or thermal instability appear to be sufficient
\citep{kim03, piontek04, piontek05, piontek07, yang07}.  Again, gravitational
instability is an obvious candidate to drive the turbulence in the weakly
star-forming portions of galaxies \citep[although see][who show that
gravitational instability is not globally important in very low-mass
dwarfs]{elmegreen15}.

A final surprising observation concerns the metallicity distributions of
galaxies.  Observations of disk galaxies show that they have slightly negative
gas-phase metallicity gradients inside $r_{25}$ \citep[e.g.,][]{vila-costas92,
  pilyugin04, henry10, balser11, ho15}, and nearly flat distributions of
metallicity outside $r_{25}$ \citep{bresolin09, bresolin12, werk11}. Dwarf
galaxies show no metallicity gradients at all \citep{croxall09}. While the
gradients in inner disks might be explicable simply via a balance between
inflow, outflow, and star formation (e.g., see \citealt{portinari00} and
\citealt{spitoni11} versus \citealt{ho15}), this is not the case for the flat
gradients in outer disks. In these strongly gas-dominated regions, the
metallicity should simply scale with the stellar mass fraction, regardless of
the presence of either inflow or outflow. While the stellar mass fraction
changes sharply with galactocentric radius outside $r_{25}$, the metallicity
does not, a finding that seems extremely difficult to explain without invoking
some form of metal redistribution. As with the previous two results,
gravitational instability seems a candidate redistribution mechanism worthy of
further exploration.

\subsection{Theoretical Studies of Gravitational Instability}

Given the likely importance of gravitational instability as a driver of galaxy
evolution, it is not surprising that there have been a number of theoretical
studies devoted to it. Some of the earliest were 1D models by \citet{lin87a,
  lin87b}, \citet{olivier91}, and \citet{ferguson01}, who argued that the
observed exponential distribution of stars in galactic disks could only be
understood if gas undergoes significant viscous transport on a timescale
comparable to the timescale over which the stellar disk forms.  These models
assumed a fixed dimensionless viscosity, but \citet{krumholz10} (generalizing
earlier work by \citet{bertin99} in the context of Keplerian disks with no
stars) showed that the viscosity and thus the rates of mass and angular momentum
transport could be computed self-consistently by balancing energy loss against
turbulence generation by gravitational instability.

Building on this work, \citep{forbes12, forbes14} argue that the observed
structure of star forming disks is a natural result of an equilibrium between
gravitational instability, accretion, and star formation. Cosmological accretion
brings gas to the disk outskirts, where the gas is in general stable to
collapse.  Over the course of a few rotation periods, gravitational instability
drives torques that tend to move gas to smaller galactocentric radii. The inward
flow of gas feeds active ongoing star formation in the inner disk, such that the
total star formation in the inner disk is ultimately modulated by the accretion
of gas at large galactocentric radii.  In these models the bulk of the disk is
at all times near $Q_{\rm total} \sim\, 1$.

Our motivation for 3D simulations is that, while the 1D models are instructive
and useful for exploring parameter space quickly, they have substantial
limitations. The gas and stars both contribute to the self-gravity of the disk,
so the degree of gravitational instability of both components must be considered
\citep{lin66, jog84a, jog84b, bertin88, wang94, romeo11}, in the form of the
\citet{toomre64} stability criterion. While this is possible in 1D models, it
requires strong assumptions about the vertical and azimuthal structure of the
disk, and about the response of each component. Moreover, material at a given
radius is only allowed to move inward or outward at one bulk velocity. Properly
modeling the full non-linear multi-component gravitational instability process
requires 3D modeling of both the gas and stars embedded in a realistic galactic
gravitational potential field.

Recently, a new class of high resolution numerical galaxy formation simulations
have arisen that begin to probe the density and resolution scales necessary to
resolve star forming clouds. These models include a variety of simplifying
assumptions, including an analytic stellar background potential, both with and
without imposed spiral arms \citep{dobbs06, tasker09, tasker11, smith14,
  dobbs15, tasker15}. Others resolve the gas to very high spatial resolution at
the cost of capturing only a fraction of a galactic dynamical time, or of
simulating a galaxy much smaller than the Milky Way \citep{bournaud10,
  renaud13}.  Still others take a similar approach to ours \citep{agertz09,
  hopkins12, fujimoto14a, fujimoto14b, behrendt15}, including runs based on the
same AGORA \citep{kim14} initial conditions that we use \citep{agertz15}.

In this paper we make use of simulations of the gas, stars, and dark matter in
idealized isolated models of star forming disk galaxies.  We focus on the
spatial and mass scale similar to the Milky Way and study three different disks
with varying gas fractions. We expand on previous work in a number of ways,

\begin{itemize}

\item Using our high numerical resolution --- $\Delta x \sim\, \unit[20]{pc}$ at
  the maximum refinement level, corresponding to a threshold density for star
  formation, $n_{\rm thresh} \sim\, \unit[50]{cm^{-3}}$ --- we are able to
  comfortably resolve the global properties of star forming regions in our disk
  models (although we do not resolve the internal properties of star forming
  clouds), allowing us to resolve an ISM with multiple distinct phases.

\item Our simulations are run for several galactic dynamical times. This long
  time baseline allows the instability to reach a saturated, fully chaotic state. As we will
  show, the onset of violent gravitational instability in our disk models is
  accompanied by substantial transient ringing behavior characteristic of the
  smooth unstructured initial conditions, and we run for long enough for this
  initial state to be forgotten.

\item We perform a detailed modeling of the Toomre $Q$ parameter, taking into
  account contributions of the gas and stars, including both the thermal and
  kinetic gas pressure, and accounting for the nonzero thickness of the gas and
  stellar disks.

\item We measure the radial and temporal variation in the rate of radial mass
  transport at high spatial and temporal resolution, allowing us to track the
  detailed dynamics of the flow of matter in the disk and measure a mean flow of
  matter through the disk.

\item The initial conditions data, full simulation code, simulation outputs, and
  analysis code are publicly available\footnote{\url{http://dx.doi.org/10.13012/J8F769GV}}.

\end{itemize}

Our simulations are quite similar to those of \citet{agertz15}. Their
simulations are based on nearly identical initial conditions but were run with a
factor of two higher spatial resolution than we use. On the other hand, we
simulate galaxies over much longer baselines (\unit[600]{Myr}
vs.~\unit[140]{Myr}), which is essential to allow the gravitational instability
to settle reach full non-linear saturation and thereby to allow an accurate
measurement of the rates of the typical mass transport rate due to gravitational
instability. \citet{agertz15} do not consider the issue of transport in their
study.

In addition, our simulations can be compared to the numerical experiments of
\citet{behrendt15}. They focused on the transition from the linear to non-linear
phase of gravitational instability in the context of an analytic treatment of
the Toomre instability. We see similar morphology in our simulations --- an
initially smooth disk that breaks up into axis-symmetric rings which then
further segment and collapse into bound clumps. Their simulations were run with
an isothermal equation of state and did not include a stellar disk and
furthermore did not consider radial mass transport.

This paper is part of a series. In the companion paper (Goldbaum et al., 2015,
in preparation) we perform simulations including stellar feedback, and compare
these to the pure-gravity simulations presented here. A previously published
paper in this series \citep{petit15} makes use of one of the suite of
simulations we describe here in order to study metal transport by gravitational
instability, and thus we will not examine that topic further in this work.

The plan for the remainder of this paper is as follows.  In
\autoref{simulations} we describe the code, subgrid physics models, and initial
conditions. In \autoref{qualitative}, we describe in general terms the time
evolution of our simulations, particularly with respect to the dynamical
evolution of the gravitational instability process. In \autoref{results} we
quantitatively measure the effect of the gravitational instability via the time
evolution of the $Q$ parameter, as well as the radial mass flux, measure the
mean mass flux rate through our simulated disks, and discuss the velocity
structure of the gaseous disk.  Finally in \autoref{discussion} we review our
findings and discuss implications for future work.

\section{Simulations}

\label{simulations}

This paper makes use of three simulations of isolated disk galaxies.  All
simulations were performed using the \texttt{Enzo} AMR hydrodynamics code
\citep{enzo14}. These simulations include self-gravity, cooling, high-order
shock-capturing hydrodynamics, and N-body dynamics. The properties of the
simulations are summarized in \autoref{simulation-table}.

\subsection{Initial Conditions}

The initial conditions were generated for the AGORA project \citep{kim14} using
the \texttt{makegalaxy} code \citep{springel05}.  Briefly, \texttt{makegalaxy}
transforms an input halo mass ($M_{200}$), stellar disk mass ($M_*$), gas
fraction ($f_g$), halo spin parameter ($\lambda$), and halo concentration
parameter ($f_c$) into particle initial conditions. These input parameters
uniquely set the rotation curve, $V_C(R)$, and radial scale length, $h_R$ via
the analytic theory of \citet{mo98}.

\begin{deluxetable}{cc}
\tabletypesize{\footnotesize}
\tablecolumns{2}
\tablewidth{0pt}
\tablecaption{Simulations}
\startdata\
  \\
  \cutinhead{Halo Parameters}
  $M_{200}$ & \unit[$1.1 \times 10^{12}$]{$M_\odot$} \\
  $V_{200}$ & \unit[150]{km s$^{-1}$} \\
  $R_{200}$ & \unit[206]{kpc} \\
  $f_c$ & 10 \\
  $\lambda$ & 0.04 \\
  \\
  \cutinhead{Disk Parameters}
  $M_*$ & \unit[$4.3 \times 10^{10}$]{$M_\odot$} \\
  $h_R$ & \unit[3.4]{kpc} \\
  $h_z$ & \unit[0.34]{kpc}
  \\
  \cutinhead{Bulge Parameters}
  $M_B$ & \unit[$4.3 \times 10^{10}$]{$M_\odot$} \\
  \\
  \cutinhead{Resolution Parameters}
  $\Delta x$ & \unit[20]{pc} \\
  $n_{\rm{thresh}}$ & \unit[50]{cm$^{-3}$} \\
  $N_{H}$ & $10^7$ \\
  $N_{D}$ & $10^7$ \\
  $N_{B}$ & $1.25 \times 10^6$ \\
  $m_{H}$ & \unit[$1.3 \times 10^5$]{$M_\odot$} \\
  $m_{D}$ & \unit[$3.4 \times 10^3$]{$M_\odot$} \\
  $m_{B}$ & \unit[$3.4 \times 10^3$]{$M_\odot$} \\
  \\
  \cutinhead{Gas fraction}
  LGF & $10\%$ \\
  Fiducial & $20\%$ \\
  HGF & $40\%$
  \enddata\

  \tablecomments{All three simulations are initialized with the same halo, disk,
    bulge, and resolution parameters.}
  ~\label{simulation-table}
\end{deluxetable}

The dark matter and stars are represented in the galaxy model using
collisionless particles and are initialized by stochastically drawing from
analytic distribution functions. The dark matter positions are initialized to
follow a \citet{hernquist90} distribution, which closely matches the more
commonly used \citet*{navarro96} fitting formula, but is more analytically
tractable. The star particle positions are initialized following
\begin{equation}
\rho_*(R, z) = \frac{M_*}{4 \pi h_z h_R^2}
{\rm sech}^2\left(\frac{z}{2 h_z}\right)\exp\left(-\frac{R}{h_R}\right)
\end{equation}
where $M_*$ is the mass of the stellar disk, $h_z$ is the
vertical scale height, $z$ is the vertical coordinate, and $R = \sqrt{x^2 +
  y^2}$ is the cylindrical radial coordinate. We consider gas fractions $f_g =
0.1$, $0.2$, and $0.4$, which we refer to as the low gas fraction (LGF),
fiducial, and high gas fraction (HGF) runs, respectively
(\autoref{simulation-table}).  Once the particle positions are computed, the
velocities are populated using a distribution function that depends only on the
local orbital energy $E$ and vertical component of the angular momentum $L_z$.
The velocity distribution function is assumed to be axissymetric, making it
strraightforward to locally solve the Jeans equation making use of the known
density distribution from above (see \citet{springel05} for more details).

In addition to the stellar disk component, we also include a bulge of mass
$M_B$.  The bulge is initialized following a Hernquist profile:
\begin{equation}
\rho_B(r) = \frac{M_B}{2\pi}\frac{b}{r{(r+h_B)}^3}
\end{equation}
where $h_B$ is a free parameter.  The bulge-to-total mass ratio,
$M_B/(M_B+M_*)$, is 0.1 in all three runs.

We include $N_H = 10^7$ halo particles, $N_D = 10^7$ particles in the stellar
disk, and $N_B = 1.25 \times 10^6$ particles in the stellar bulge. All particles
in each population have uniform masses. For the halo population, the mass of
each particle is $m_H = 1.3 \times 10^5 M_\odot$. For the bulge and disk
population the particle masses are $m_D = m_B = 3.4 \times 10^3 M_\odot$.

Since \texttt{makegalaxy} produces initial conditions formatted for the
\texttt{gadget} smoothed particle hydroynamics (SPH) code, special care must be
taken to initialize the gas onto \texttt{Enzo}'s AMR grid structure.  Rather
than interpolating from the initial conditions for the SPH particles, we instead
initialize the gas density following an analytic density profile:
\begin{equation}
  \rho(R, z) = \frac{M_G}{4 \pi h_R^2  h_z}
  \exp\left(\frac{-R}{h_R}\right)\exp\left(\frac{-|z|}{h_z}\right)
\end{equation}
where $M_G = f_g M_*$ is the mass of the gaseous disk. The gas velocities are
initially axisymmetric, with the radial profile set according to the circular
velocity curve written out by \texttt{makegalaxy}.

This departs somewhat from the procedure used by \texttt{makegalaxy} to generate
SPH initial conditions, where the gas density distribution is assumed to be
exponential in the radial direction but the vertical extent is determined via an
iterative relaxation process to ensure the gas disk is initially in
equilibrium.  This relaxation process assumes that the gas pressure is moderated
by a subgrid effective equation of state model, which we do not employ in this
study.  In practice, the gas in our simulations initially experiences a phase of
violent collapse and relaxes into a quasi-equilbrium state after the
gravitational instability has fully developed.  This means the precise state of
the gas in the initial conditions is not terribly important --- the simulation
``forgets'' the gas initial conditions. Furthermore, this choice also allows us
to make an apples-to-apples comparison with future isolated galaxy simulations
that make use of the public AGORA isolated galaxy initial conditions.

\subsection{N-body dynamics, Hydrodynamics}

Rather than making use of an analytic dark matter and stellar potential
\citep[see, e.g.,][]{dobbs06, tasker09}, we employ a live dark matter halo and
stellar disk. This allows us to follow the active response of the stars and dark
matter to the collapse of the gaseous disk.  The particle dynamics are
implemented in \texttt{Enzo} using a standard particle-mesh scheme
\citep{hockney88}.  Particle positions and velocities are updated according to
the local gravitational acceleration using a drift-kick-drift scheme.

Hydrodynamics are captured using the piecewise parabolic method (PPM)
\citep{colella84}. PPM hydrodynamics allows us to accurately capture strong
shocks in a few computational zones while also maintaining second-order spatial
accuracy. While the PPM method is formally second-order accurate in space and
time, \texttt{Enzo}'s adaptive timestepping scheme is only first-order accurate, so these
calculations are only first order accurate in time. In addition, after
experiencing intermittent instability using second-order interpolation, we opted
to degrade to first-order interpolation at AMR level boundaries. This means that
we are formally first-order accurate in space, although these inaccuracies
should only show up at level boundaries.

Since the \unit[$\sim\,$200]{km s$^{-1}$} circular velocity of the galaxy
necessitates strongly supersonic flows in the galactic disk, we make use of the
dual energy formalism implemented in the \texttt{Enzo} code \citep{bryan95}. In
standard PPM hydrodynamics, the internal energy is not tracked --- instead it is
derived from the total energy and velocity.  When the kinetic energy is much
larger than the internal energy, this can lead to spurious temperature
fluctuations due to floating point round-off error.  To avoid this,
\texttt{Enzo} tracks a separate internal energy field that in effect provides
extended precision to the total energy field. This allows us to safely resolve
the thermal physics of the gas in a global simulation of a disk galaxy without
worrying about spurious temperature fluctutions due to advection errors.

\subsection{Initial Grid Structure and Refinement Criteria}

Our simulations make full use of the AMR capabilities of the \texttt{Enzo}
code. Our refinement strategy is to focus computational effort on the highest
refinement level, while still adequately resolving the dark matter halo so our
estimate of the gravitational potential in the neighborhood of the galactic disk
is accurate.  Since the dark matter halo is spatially extended compared to the
disk, with the furthest dark matter particles living \unit[500]{kpc} away from
the center of the galaxy, this necessitates a hierarchically nested AMR
structure.

To fully encompass the dark matter halo, we employ a cubic simulation box with a
width of \unit[1.3]{Mpc}, and resolve the root grid with $64^3$ cells.  In
addition to the static root grid, we impose 5 additional levels of statically
refined regions, enclosing volumes that are successively smaller by a factor of
8.

In addition to the static refinement, we allow for an additional five levels of
adaptive refinement.  To keep the particles properly resolved at all times, we
refine a cell if the total mass in particles within the cell exceeds \unit[$1.7
\times 10^6$]{M$_\odot$}, or approximately 10 halo particles.  This choice
produces nested grid hierarchies in the region of the simulation dominated by
the dark matter halo where the grid hierarchy is determined adaptively.

To keep the gaseous disk resolved at all times, we refine a cell if the total
mass in gas within the cell exceeds \unit[$2.2 \times 10^4$]{M$_\odot$}. Lastly,
to avoid artificial fragmentation, we ensure the Jeans length,
$\lambda_J = \sqrt{(\pi c_s^2)/ (G \rho)}$ is locally resolved at all times by
at least 32 cells, comfortably satisfying the \citet{truelove98} criterion. To
keep the Jeans length resolved after collapse has reached the maximum refinement
level, we employ a pressure floor such that the Jeans length is resolved by at
least 4 cells on the maximum refinement level. To avoid contaminating the
hydrodynamics in situations where we can use more resolution, the pressure floor
is only employed for cells on the maximum refinement level. The pressure floor
is implemented as a source of ``extra'' pressure above the thermal pressure --
an improvement over the Jeans stabilization routines in the current stable
version of \texttt{Enzo} where the pressure floor is implemented by increasing
the temperature.  The new pressure floor implementation will be included in the
next stable release of \texttt{Enzo}.

The Enzo AMR hierarchy is fully adaptive in space and time, so regions that no
longer satisfy the refinement criteria described above will no longer be
refined. We do not include an explicit de-refinement criteria, instead allowing
the refinement criteria described above to fully control the AMR hierarchy.

The particle refinement criteria we use are not as stringent as the gas
refinement criteria. If the gas were not present in the simulation, the star
refinement criteria alone would cause the stellar disk to be refine to AMR level
7 (\unit[160]{pc}) resolution in the outskirts of the disk and AMR level 8
(\unit[80]{pc} resolution in the inner disk and bulge.  Since the combined
contribution of the gas and stars determine the final resolution of the stars,
in general the stellar disk is resolved to $\unit[20-40]{pc}$ resolution. In
particular, the resolution is higher near dense clumps and filamentary structure
in the gas.  Early in the simulation, before the gas has had a chance to be
consumed by star formation, the resolution is also generally higher.

Near the beginning of all three simulations, there are $3 \times 10^7$
computational zones in total, with about $1.5 \times 10^7$ zones on the maximum
refinement level.  Towards the end of the fiducial and high gas fraction
simulations, after much of the gas disk has been consumed the number of zones on
the maximum refinement level decreases to $\sim 5 \times 10^6$ zones. We
attribute this decline to the collapse of gas into dense clumps and the
consumption of gas by star formation, leaving much of the volume occupied by
relatively rarefied gas.

\subsection{Star Formation}

Since the gaseous disks in our simulations are unstable to fragmentation and
collapse, it is necessary to include a subgrid model that converts dense,
collapsing gas into newly formed star particles. Rather than using the built-in
``standard'' star formation prescription in the \texttt{Enzo} code
\citep{enzo14, cen92}, which is tuned for lower-resolution cosmological
simulations, we make use of a new star formation prescription. To ease
comparison with future papers using the AGORA initial conditions, the
prescription is based on the suggested star formation model for the AGORA
project \citep{kim14}.

Briefly, the model assumes that the star formation rate density in any cell is a
function only of the gas density in that cell, according to the following
formula:
\begin{equation}
\label{sf_law}
\frac{d \rho_*}{dt} = \left\{
     \begin{array}{lr}
       f_* \frac{\rho}{t_{\rm ff}} & : \rho > \rho_{\rm thresh}\\
       0 & : \rho \le \rho_{\rm thresh}
     \end{array}
   \right.
\end{equation}
where $f_*$ is the star formation efficiency,
$t_{\rm ff} = \sqrt{3 \pi/32 G \rho}$ is the local dynamical time, and
$\rho_{\rm thresh} = \mu\, m_h\, n_{\rm thresh}$ is the threshold density for
star formation. Here $\mu$ is the fixed mean molecular weight, $m_h$ is the mass
of a hydrogen atom, and $n_{\rm thresh} = \unit[50]{cm^{-3}}$ is the threshold
number density. Note that since we do not advect species fractions for hydrogen
and helium ionization states, $\mu = 1.4$ is constant over the full simulation
box. The value we choose is typical of the bulk of the atomic ISM.\@ In all
of our simulations, we use $f_* = 1 \%$, in accordance with the observed low
star formation efficiency universally observed in star forming regions at a
range of density and size scales \citep{krumholz07, krumholz12}.

The star formation threshold is chosen such that the Jeans length for gas on the
maximum refinement level remains resolved by at least 4 cells until the
refinement reaches the maximum level, assuming a temperature floor of
\unit[100]{K}, approximately the minimum temperature we see in our
simulations. Once gas reaches the maximum allowed refinement level, the gas is
no longer well resolved, and we allow the gas to convert itself into stars and
apply a pressure floor to prevent artificial fragmentation. Since we only barely
resolve the typical densities for the formation of molecular gas, we do not
employ a subgrid model to track the molecular gas fraction, instead assuming
that gas above the threshold density is fully molecular. This follows
\citet{guedes11}, who found that the precise value chosen for
$\rho_{\rm thresh}$ was not important for determining the star forming
properties of a disk galaxy in a cosmological zoom-in simulation, so long as the
threshold density is high enough that star formation occurs primarily in dense
clumps of gas rather than in the bulk of the ISM\@.

Practically speaking, \autoref{sf_law} cannot be solved by spawning star
particles at each timestep for all cells with densities above the threshold
density. Given a typical timestep of \unit[1000]{yr}, and assuming a cell
containing gas at the threshold density, we would expect to be spawning star
particles with masses of only \unit[.01]{M$_\odot$}.  Rather than spawning star
particles in every cell that exceeds the density threshold, we instead impose a
minimum star particle mass, and form stars stochastically.  In our scheme, a
cell will form a star particle of mass $m_{\rm sf} = \unit[300]{M_\odot}$ with
probability
\begin{equation}
P_* = \left\{
  \begin{array}{lr}
    f_* \frac{\rho \Delta x^3}{m_{\rm sf}} \frac{dt}{t_{\rm ff}} &: \rho >
    \rho_{\rm thresh}\\
    0 &: \rho \le \rho_{\rm thresh}
    \end{array}
\right.
\label{stochastic_sf_law}
\end{equation}
where $dt$ is the simulation timestep on the maximum refinement level. By
stochastically creating according to \autoref{stochastic_sf_law}, the
time-averaged value of the star formation rate in each cell is still given by
\autoref{sf_law}, while creating a reasonable number of particles.

\begin{figure}
\epsscale{1.1}
\plotone{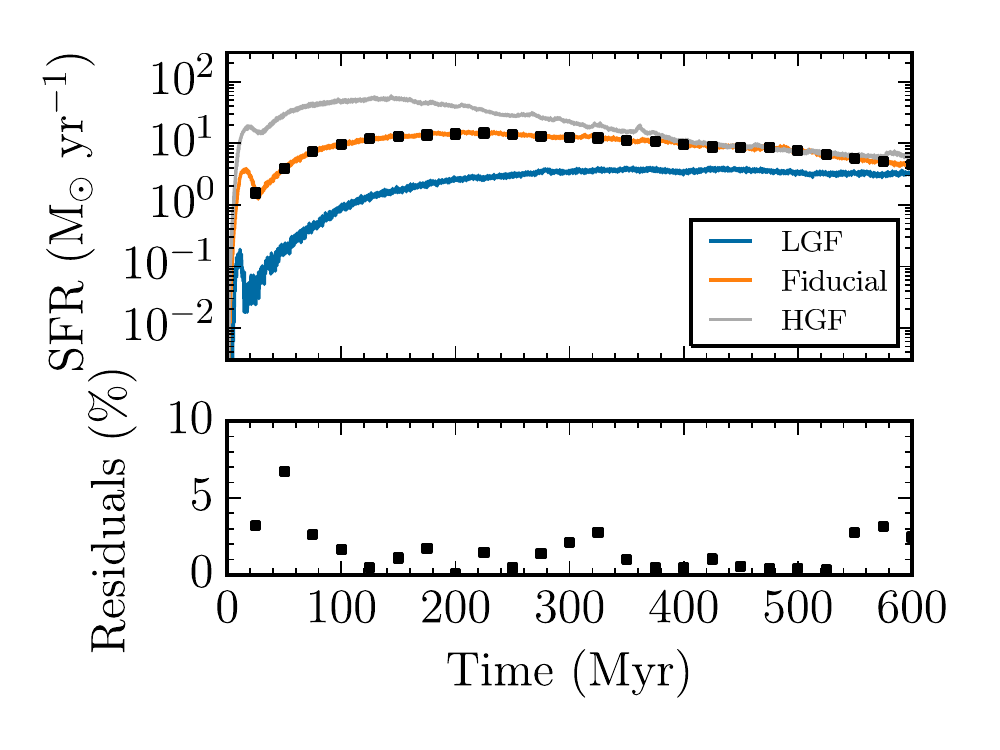}
\caption{The observed (solid lines) star formation rate as a function of time
  for all three simulations.  The black squares were calculated by directly
  measuring the expected star formation rate in the fiducial simulation
  according to \autoref{sf_law}. The blue, orange, and grey lines were
  calculated by binning the star particles that are dynamically formed in the
  simulation by their formation times. In the bottom panel we plot the residual
  between the black squared and orange line in the top panel. We find good
  agreement between the predicted and measured star formation rate.  \\}
\label{sfr_figure}
\end{figure}

In \autoref{sfr_figure}, we compare the star formation rate inferred from the
ages of star particles present at the end of each simulation with the expected
star formation rate measured by evaluating \autoref{sf_law} at \unit[25]{Myr}
intervals.  We see that there is good agreement between the measured and
predicted star formation rate, with a maximum deviation of $\sim\,$7 \%.  The
implementation of this star formation algorithm has been made publicly available
and will included in the next stable release of \texttt{Enzo}.

\subsection{Heating and Cooling}

To model the thermal physics of the gas in our simulations, we make use of the
\texttt{Grackle} cooling library\footnote{\url{https://grackle.readthedocs.org}}
\citep{enzo14, kim14}. The \texttt{Grackle} cooling and heating routines were
adapted from the chemical and thermal physics implementation included in the
\texttt{Enzo} code. Relative to \texttt{Enzo}'s implementation of thermal
physics, \texttt{Grackle} adds novel capabilities that we make use of in this
study. In addition, this choice eases comparison with future simulations
performed in other codes as well as simulations performed for the AGORA
comparison.

\texttt{Grackle} includes a primordial cooling routine based on tabulated
cooling rates as a function of density and temperature, avoiding the need to
evolve separate fields for each hydrogen and helium ionization state. In
addition to cooling by primordial species, we also include cooling due to metal
line emission. These cooling rates are inferred from tabulated rates output by
the \texttt{CLOUDY} code.

The disk is initialized to include a metal color field that follows the initial
gas distribution. All gas zones in the disk are initialized with a solar metal
fraction.  The metal field is passively advected along with the gas. We include
a dynamical metal color rather than assuming a fixed metal fraction in
anticipation of modeling the production of metals in simulations with supernova
feedback.

To properly model the thermal physics in a Milky Way-like ISM, we include a
prescription for heating due to electrons released from dust grains by the
photoelectric effect. This is implemented as a constant heating rate of
$8.5\times10^{-26}\ \rm{erg}\ \rm{s}^{-1}$ per Hydrogen atom applied uniformly
throughout the simulation box for gas below $10^{4.3}$ K \citep{tasker11}.  This
rate is chosen to match the expected heating rate assuming a UV background
consistent with the solar neighborhood value \citep{draine11}.

\subsection{Analysis}

In order to analyze our simulations, we construct an extensive post-processing
pipeline using a combination of the \texttt{yt} toolkit \citep{turk11} and a set
of custom python analysis scripts that we have made publicly available on
Bitbucket\footnote{\url{https://bitbucket.org/ngoldbaum/galaxy_analysis}}.
Details of the analysis pipeline, and how we reconstruct various quantities of
interest from the raw simulation outputs, are described in \autoref{analysis}.
In particular, \autoref{analysis} gives our formal definitions of quantities
that are non-trivial to calculate, including velocity dispersions
(\autoref{velocity_dispersion}), effective sound speeds
(\autoref{ceff}), and Toomre $Q$ parameters (\autoref{toomre_q}).

\section{Qualitative Outcome}

\label{qualitative}

\begin{figure*}
\epsscale{1.15}
\plotone{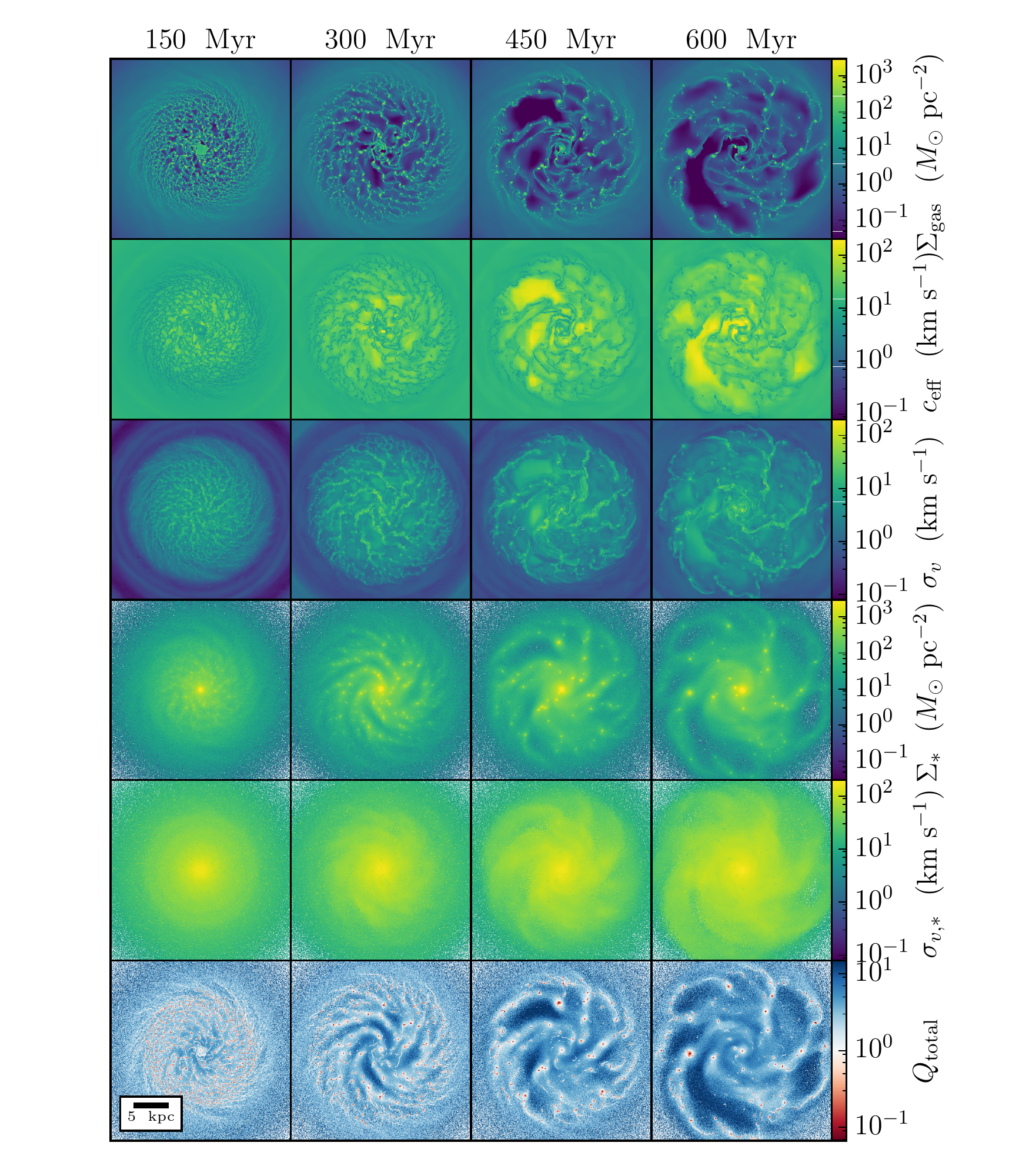}
\caption{The time evolution of the gas and stars in the fiducial simulation. The
  quantities shown are, from top to bottom, gas surface density, effective sound
  speed of the gas (including thermal and turbulent contributions), gas velocity
  dispersion, stellar surface density, stellar velocity dispersion, and total
  (gas plus stars) Toomre $Q$; formal definitions for all quantities are given
  in \autoref{analysis}.  The simulation time for each column is indicated at
  top and the spatial scale is indicated by the scale bar at the bottom
  left. Each panel displays a region \unit[25]{kpc} across centered on the
  galaxy.}
\label{evolve_summary}
\end{figure*}

\begin{figure*}
\epsscale{0.9}
\plotone{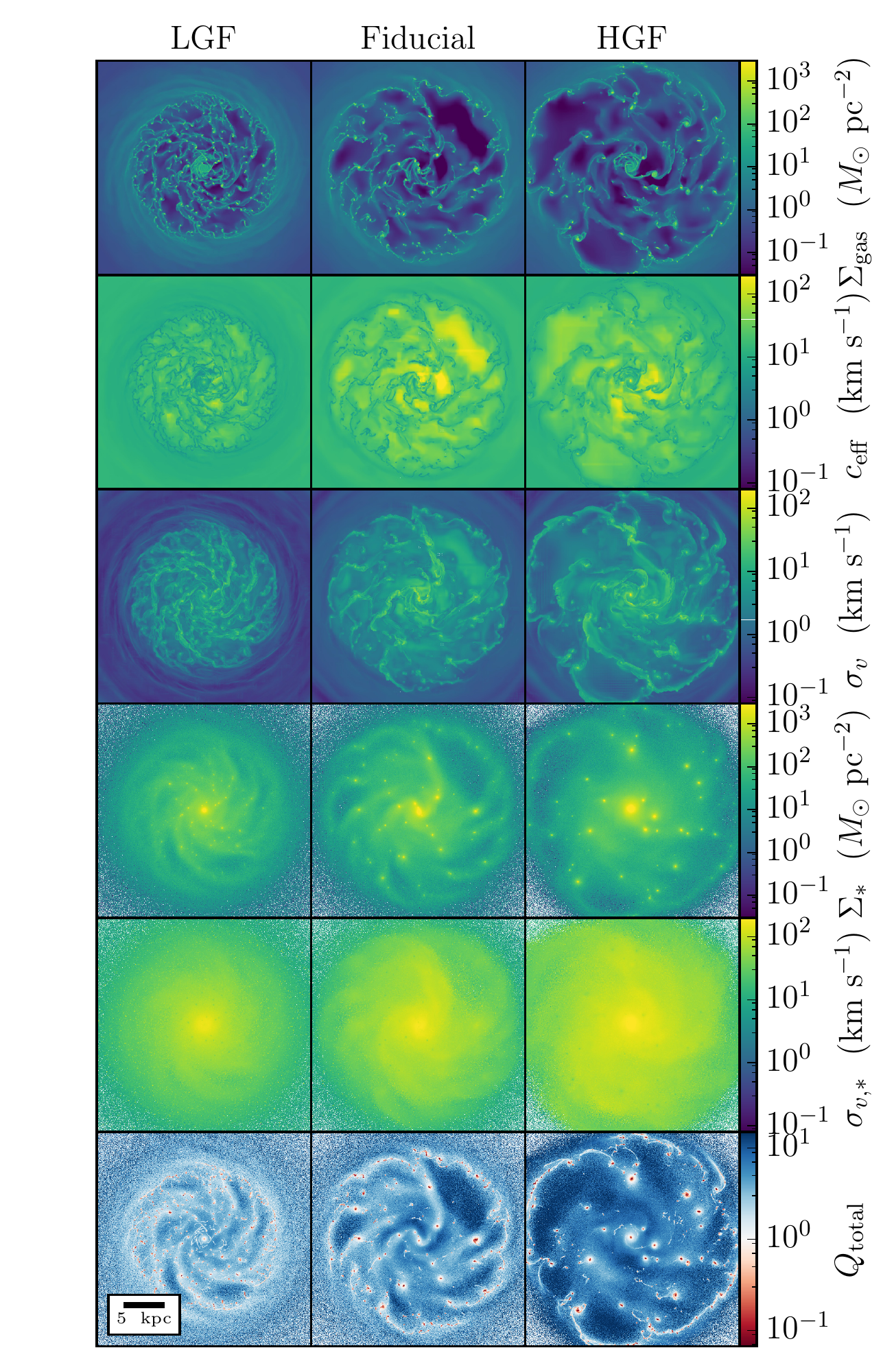}
\caption{Same as \autoref{evolve_summary} but for each of the
  three different simulations at $T=\unit[500]{Myr}$.}
\label{sims_summary}
\end{figure*}

\begin{figure}
\epsscale{1.1}
\plotone{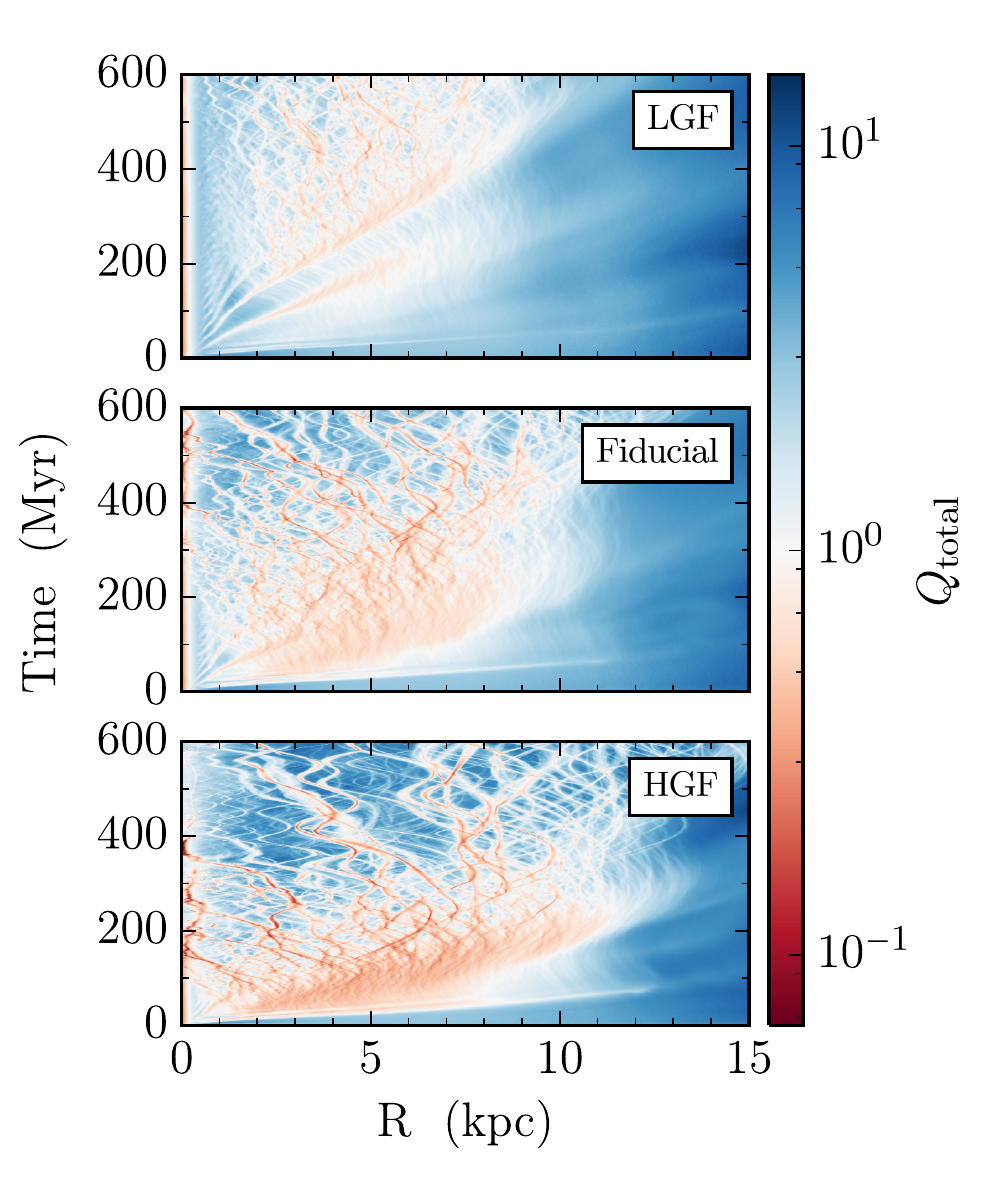}
\caption{The time evolution of the azimuthally averaged Toomre $Q$ parameter in
  each of our simulations (see \autoref{romeoq}). This estimate includes
  the combined contribution of both the gas and stars. The color scale is chosen
  such that regions that are gravitationally unstable are colored red, regions
  that are marginally stable are colored white, and regions that are stable are
  colored blue.\\}
\label{q_evolution}
\end{figure}

\begin{figure*}
\plotone{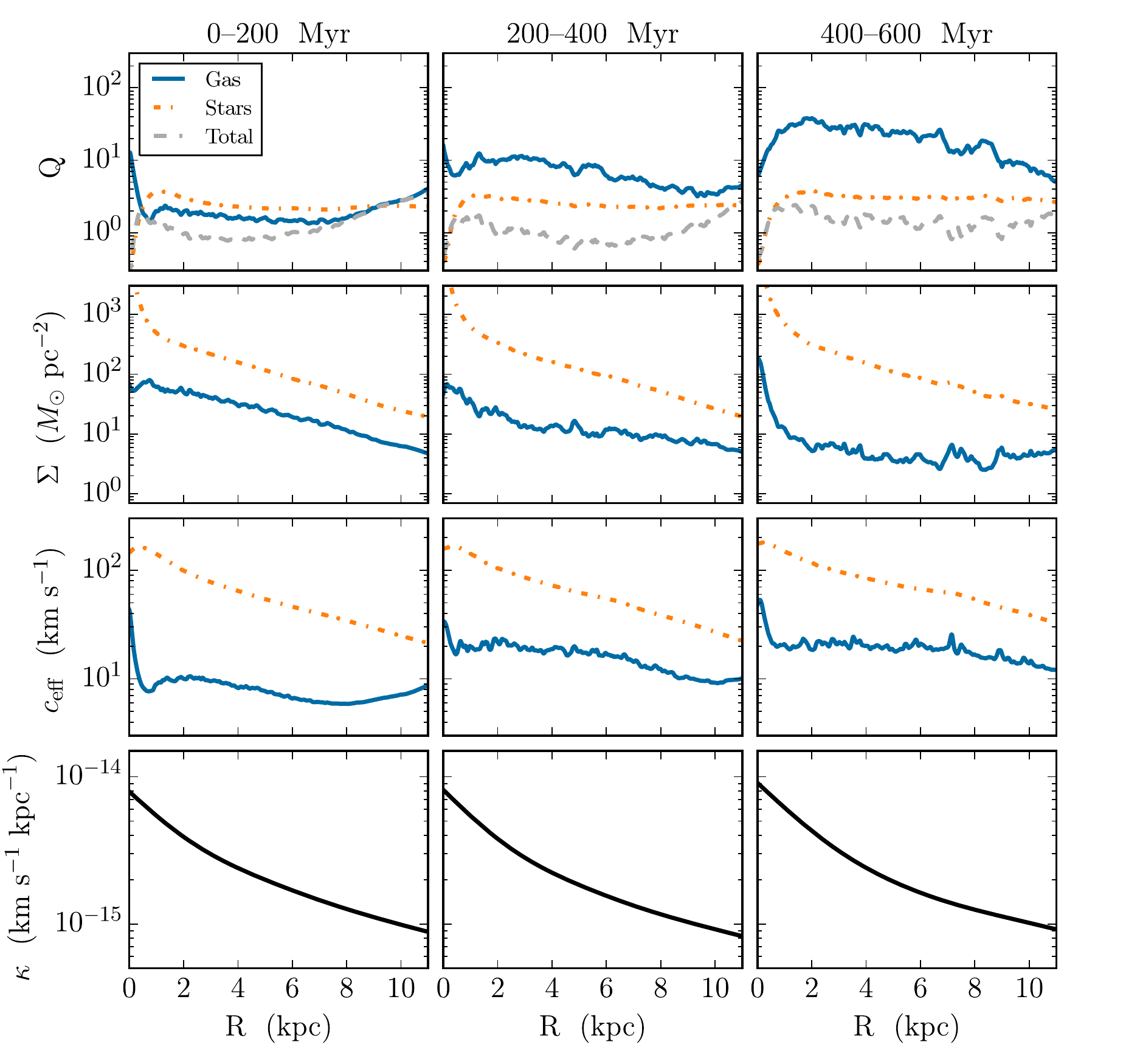}
\caption{Time averages of the Toomre Q parameter (top row), the surface density
  (second row), and effective sound speed (third row) for the gas and stars
  averaged over three different periods (indicated at top) in the fiducial
  simulation. The bottom row shows the local epicyclic frequency $\kappa$, which
  does not vary significantly over the course of the simulation. The evolution
  in $Q_{\rm total}$ is primarily driven the evolution in $Q_{\rm gas}$, which
  in turn is driven by depletion in the gas supply and and increase in the gas
  velocity dispersion.\\}
\label{q_summary}
\end{figure*}

We present snapshots of the fiducial simulation at four times in
\autoref{evolve_summary} and snapshots of all three simulations at a fixed
time in \autoref{sims_summary}.  Each figure displays the gas and stellar
surface density and effective sound speed as well as the combined Toomre $Q$
parameter.

The dynamics of all three simulations are similar. The initially smooth gaseous
disk quickly cools from the initial temperature of \unit[$10^4$]{K} to
\unit[200--300]{K}. The initially thermally supported disk proceeds to collapse
from vertically over the course of the next \unit[20--50]{Myr}.  Denser regions
in the center of the disk collapse first, followed by less dense regions further
out in the disk, with the precise collapse time determined by the initial gas
surface density at any given radius. Once the gas has collapsed in the vertical
direction, the disk remains very thin for the rest of the simulation, with the
bulk of the gas only one or two cells (\unit[20--40]{pc}) away from the
midplane. Our disks are thus not well resolved in the vertical direction.

Once the gas has collapsed vertically, a combination of shear and self-gravity
shepherds the gas into filaments which in turn collapse into isolated
gravitationally bound clouds.  Since these simulations do not include feedback,
these clouds survive more or less permanently, only disappearing if they exhaust
their gas supply by converting gas into stars, or merge with one another. As the
gas in the gravitationally bound clouds is converted into stars, massive star
clusters form inside the clouds. Once the initial period of collapse and
fragmentation has passed, both the gas and stars spontaneously align into clear
spiral arms.  Later, as an increasing fraction of both the gas and stars
collects inside the gravitationally bound clumps, the spiral structure becomes
less clearly defined. These long-lived dense clouds are very similar to those
seen by \citet{hopkins12} in their simulations with no feedback.

Gas in the innermost regions is stabilized against fragmentation by the presence
of the bulge. Rather than forming gravitationally bound clouds, the gas in the
bulge quickly stabilizes into a thin disk.  Since the disk is above our density
threshold for star formation, the central disk disappears as it converts itself
into newly formed stars. We caution that star formation in the bulge might be an
artifact of our \unit[20]{pc} resolution, since we do not see any fragmentation
there and the bulk of the gas in the bulge is at or above our star formation
threshold density.

The combination of gravitational collapse and shear induces significant
turbulent motion in the gas, with typical velocity dispersions of
\unit[10--20]{km s$^{-1}$} in the portion of the disk that participates in
gravitational instability ($R \lesssim \unit[15]{kpc}$). The hot, low density
gas in the interarm regions exhibits slightly higher turbulent velocity
dispersions than the dense gas, but with much lower typical Mach numbers. In the
interarm region, the effective sound speed is typically
$\unit[100-200]{km s^{-1}}$ and is dominated by the thermal sound speed, while
in the dense gas the turbulent velocity dispersion dominates the effective sound
speed. Thermal sound speeds are high in the interarm regions because
these are heated by spiral shocks, and their low densities leave them unable
to cool effectively afterwards. In contrast, the denser arm gas quickly cools 
back to its equilibrium temperature.
The stellar orbits do not experience significant heating or dissipation,
retaining the \unit[$\sim$\,40]{km s$^{-1}$} velocity dispersion present in the
initial conditions. We also note that the kinetic energy content of the interarm
gas is negligible compared to the gas in dense clumps, and does not contribute
significantly to the kinetic energy budget of the disk.

The Toomre $Q$ parameter tends to increase in time. In our initial conditions
for all three simulations, $Q \sim\, 1$ throughout the disk. As soon as the gas
is able to cool and begins to collapse, $Q \ll 1$. As collapse proceeds, the
typical surface density decreases and the degree of turbulence increases,
leading to $Q$ steadily increasing in time.  Regions in which gravitationally
bound clumps develop consistently exhibit $Q \ll 1$, even at late stages, but,
as we show below, azimuthal averages of $Q$ are always greater than one. The
locally low values of $Q$ within the bound clumps can mostly be attributed to
very high gas and stellar surface densities in these regions. We do not see any
significant increase in the stellar velocity dispersion in these regions, and we
see only a modest increase in the gas effective sound speed. See
\autoref{gravitational_instability} below for more discussion of how $Q$ evolves
with time.

The stellar and gas dynamics in all three simulations are similar, but the
detailed dynamics do vary somewhat as a function of initial gas fraction.
Increasing the gas fraction leads to quicker collapse. Decreasing the gas
fraction produces a slower collapse in which the gas spends a longer amount of
time in filaments rather than dense clumps. Spiral arms are more prominent in
the simulation with low gas fraction, presumably because they do not have enough
time to be disrupted by the formation of gravitationally bound clouds and
accompanying star clusters.  The star forming portion of the disk is also more
compact, since regions at the outskirts of the disk are no longer dense enough
to cool and collapse.

The degree of gravitational instability displays a weak dependence on the
gas fraction. Since the high gas fraction run is able to process a much larger
fraction of the gas present in the initial conditions, it reaches a higher
typical $Q_{\rm total}$ value due the lower typical gas surface densities in the
depleted gas disk. The low gas fraction run exhibits less variation, with
$Q_{\rm total} \sim\,$ 1--3 throughout the disk, though this may merely reflect
that $Q_{\rm total}$ rises more slowly with smaller initial gas fraction, and
that we have run only for a few orbits. In any case, in all three cases, the disks are
formally stable according to classical Toomre analysis, with $Q_{\rm total}
\gtrsim 1$. We consider this state to be the final, saturated result of gravitational
instability unrestrained by any form of feedback. We make the notion of what
constitutes such a steady state in our simulations more quantitative in
\autoref{ssec:transport}, where we discuss mass transport through the disk.

A robust feature of all of our simulations is that gravitationally bound clouds
form and then are unable to be destroyed by any mechanism besides gas
exhaustion. This leads to the formation of large star clusters composed of stars
that formed dynamically in the simulations. These star clusters in turn become a
significant contribution to the mass distribution in the midplane, creating
substantial streaming motions and departures from a smooth axisymmetric rotation
curve. In the companion paper (Goldbaum et al., 2015, in preparation) we show
that feedback capable of disrupting these complexes is necessary to form
realistic smooth disks and prevent the formation of unrealistically massive star
clusters.

\section{Results}

\label{results}

\subsection{Gravitational Instability}
\label{gravitational_instability}

Since our simulated galaxies do not include a prescription for star formation
feedback, the primary driver for the dynamical evolution of our model galaxies
is gravitational instability. While the initial conditions for our simulations
are formally stable, cooling allows the gas to quickly lose hydrostatic support,
leading to catastrophic collapse in the vertical direction. In addition, our
shearing self-gravitating disks are susceptible to the \citet{toomre64}
instability.

This story can be inferred by inspecting \autoref{q_evolution}, where we plot
the time evolution of the azimuthal average of $Q_{\rm total}$. The initially
stable disk quickly becomes unstable (e.g., the regions that show up in red in
the bottom half of each subplot). The instability leads to a radially expanding
wave of collapsing gas. Soon after, the gas collects in gravitationally bound
clumps, which proceed to migrate through the disk, both radially inward and
outward. The regions inside the gravitationally bound clouds are formally
unstable according to a local Toomre analysis due to their very high surface
densities. The interclump regions reach relatively high values of $Q_{\rm
  total}$ so these regions are formally stable to collapse.

This process plays out in all three simulations, albeit with varying collapse
speeds and degrees of violence. The high gas fraction case initially develops
filaments, but by $\sim 200$ Myr of evolution these have broken up into
giant clumps everywhere in the disk. As we lower the gas fraction, the transition
to the clump-dominated phase takes longer, and the clumps themselves
become smaller. Nonetheless, all three simulations reach similar clump-dominated
regimes, which appears to represent the fully-saturated state of the
gravitational instability.

We can see what is driving the evolution of $Q_{\rm total}$ by inspecting
\autoref{q_summary}. We show the evolution in $Q_{\rm total}$ (top row) along
with the quantities that determine $Q_{\rm total}$: the gas and stellar
surface density and effective sound speed, as well as the epicyclic frequency.
We see that both the epicyclic frequency and the stellar surface density and
velocity dispersion show little variation over the course of the simulations.

On the other hand, the gas surface density and effective sound speed show
significant variation. The effective sound speed tends to increase in time and
the gas surface density tends to decrease. Both of these effects drive a secular
increase in $Q_{\rm gas}$ over the course of the simulation. In turn, this
increase in $Q_{\rm gas}$ leads to the secular increase in $Q_{\rm total}$
Eventually, once the gas supply is exhausted, $Q_{\rm gas} \gg 1$, so
$Q_{\rm total}$ approaches $Q_*$.

\subsection{Mass Transport}
\label{ssec:transport}

Here we examine the radial flow of material in the disks of our simulated galaxy
models. We measure if there is any net flow of material either from the galactic
center outward or from the outskirts inward. In addition, we examine the
detailed radial and time dependence in the gas mass flux.

\begin{figure} \epsscale{1.1} \plotone{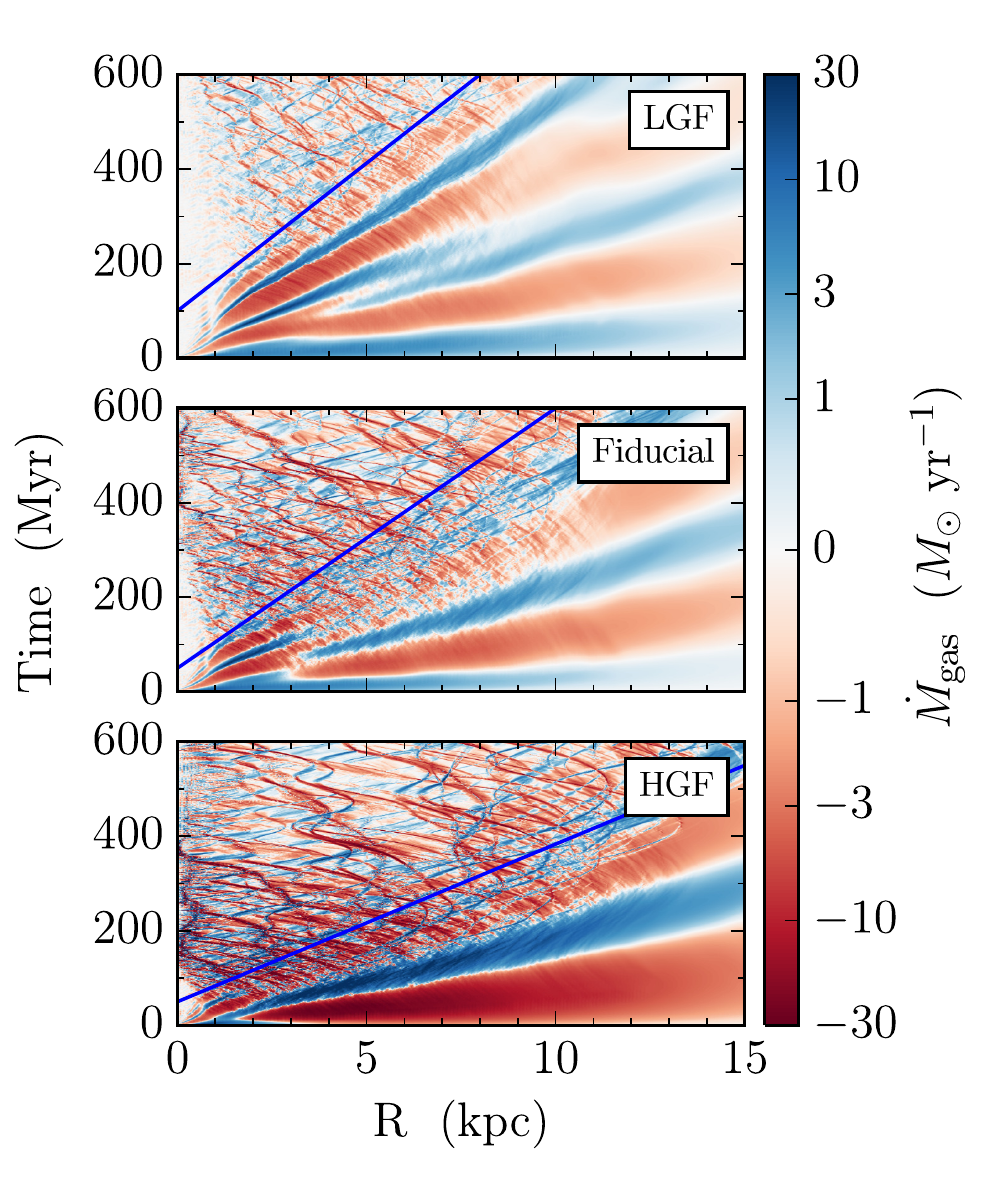}
\caption{The gas mass flux as a function of radius and time for each of our
  simulations. We define $\dot{M}_{\rm gas}$ such that positive values
  correspond to outward radial flow while negative values correspond to inward
  radial flow. The blue line indicates where we expect the disk has fully
  transitioned to an equilibrium condition, ``forgetting'' about the initial
  transients. Data below the blue line is excluded from the time averages
  plotted in \autoref{mass_flux_summary}.}
\label{radius_time_mass_flux}
\end{figure}

If \autoref{radius_time_mass_flux} we present our measurements of the gas mass
flux as a function of radius and time. All three simulations exhibit similar
overall behavior. Initially, the models exhibit significant ringing as the disks
initially collapse and settle down. In this stage the mass flux is dominated by
rings of material experiencing alternating bands of inward and outward flow.
After the initial collapse phase, the gas collects in gravitationally bound
clouds. For the rest of the simulation, the mass flux rate is primarily
determined by the inward and outward flow of the gravitationally bound
clumps. As spiral arms develop, the gravitational potential in the disk begins
to develop non-axisymmetric components that tend to drive the gas clumps both
radially inward and outward.

Along with the detailed variation in the mass flux, we would also like to know
if there is any net mass flux once the disk has settled down into a
quasi-equilibrium state. To answer this question we make use of the time
averaging algorithm described in \autoref{time_average_section}. In this way we
only consider the mass fluxes measured above the blue lines in
\autoref{radius_time_mass_flux}, which delineate the approximate time at which
the disk has settled into a steady state.

\begin{figure} \epsscale{1.0} \plotone{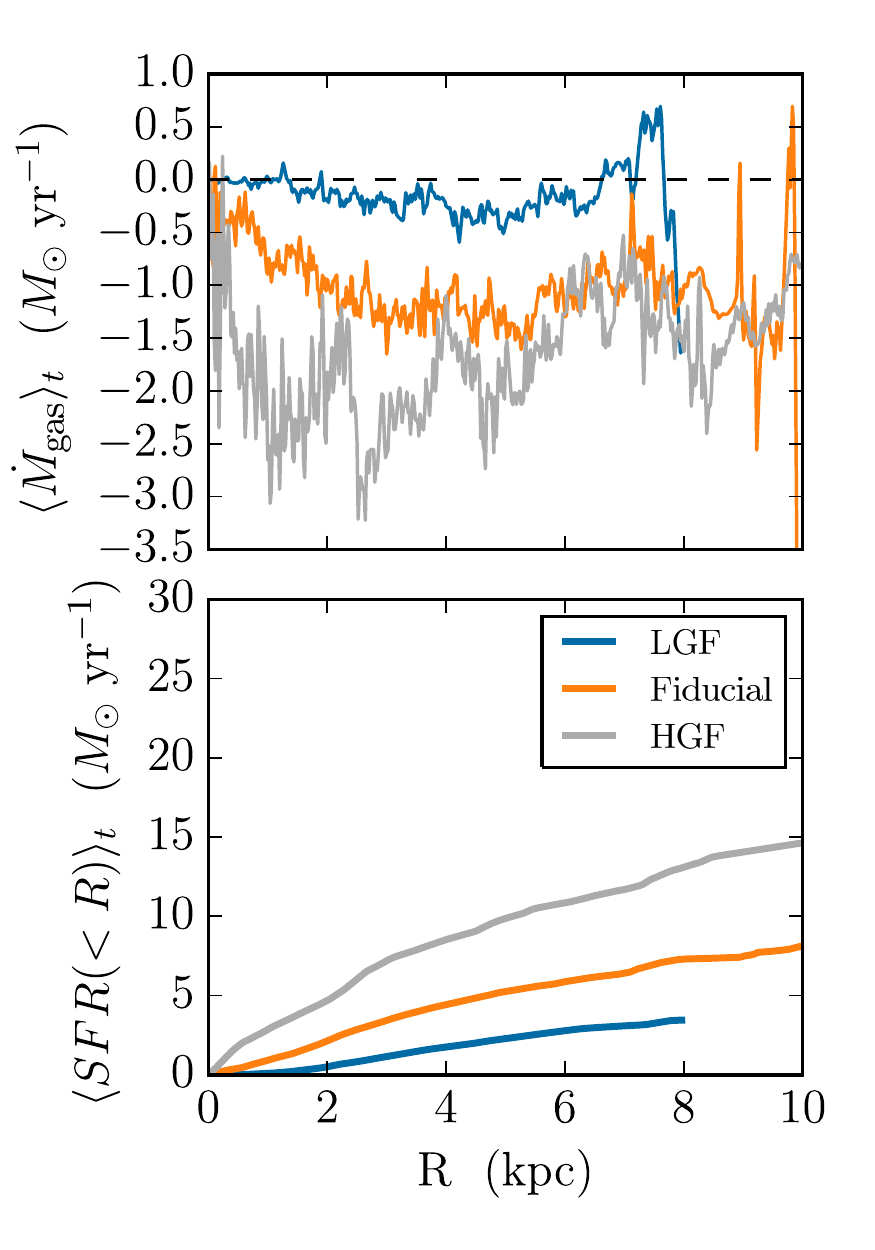}
  \caption{The time averaged mass flux (top panel) and radially cumulative star
    formation rate (bottom panel) as a function of galactocentric radius for all
    three simulations.  For the top panel, positive values correspond to outward
    radial flow while negative values correspond to inward radial flow. In all
    cases there is a significant net inward flow of material toward the galactic
    center in the inner galaxy.}.
\label{mass_flux_summary}
\end{figure}

This averaging procedure results in the time-averaged mass-flux measurements in
the top panel of \autoref{mass_flux_summary}. For each simulation, we plot the
time-averaged mass flux as a function of radius.  We find that there is a net
inward flux of gas at most radii for all three model galaxies. The flux scales
very roughly with the gas fraction, with a typical mass flux of $-0.3$, $-1.0$,
and $-2.0$ \unit{$M_\odot$ yr$^{-1}$} mass flux in the low gas fraction,
fiducial, and high gas fraction cases, respectively.

Although the flow is inward at most radii in the disk, we note that the outer
edges of the disks instead show an outward mass flux. (In
\autoref{mass_flux_summary}, this feature is outside the plotted region for the
HGF case, but is visible for the Fiducial and LGF cases.) Comparing with
\autoref{q_evolution}, we see that this outward mass flux appears roughly at the
boundary between the gravitationally unstable region and the still-stable outer
region of the disk. Such a reversal of the mass flow direction at the transition
between the stable and unstable regions of a gravitationally unstable disk is
consistent with the predictions of \citet{forbes14}. However, we caution that
these outer regions have been averaged over for the shortest times, and so the
exact value of the typical outward flux is uncertain.

We might expect that the inward flow of gas is sufficient to supply the star
formation in the inner regions of our model galaxies. To see if this is the
case, we plot the time-average of the radially-accumulated star formation rate
in the bottom panel of \autoref{mass_flux_summary}. The mass flux in the top
panel is sufficient to supply the star formation in the bottom panel only if the
mass flux is greater than the cumulative star formation rate at any given
radius. Due to the very high star formation rates in these simulations with no
feedback (c.f.\ \autoref{sfr_figure}), we find that the inward flow of gas is
insufficient to fuel the star formation rate in these simulations.

However, we note that this is a result of the unphysically high star formation
rates that our simulations exhibit due to the lack of star formation
feedback. If we instead consider the star formation rate that would be expected
given the gas surface density distribution combined with observed star formation
rates \citep[e.g.,][]{leroy13}, we reach the opposite conclusion: our inflow
rates are sufficient to fuel star formation at observed levels. Indeed, our
fiducial, Milky Way-like simulation produces mass transport at a rate of
$\sim\, 1$ $M_\odot$ yr$^{-1}$, which is roughly the observed star formation
rate in the Milky Way \citep[e.g.,][]{chomiuk11}.  The question of whether
inflows and star formation can be matched simultaneously in a simulation
including feedback we defer to the companion paper.

\subsection{Gas Velocity Structure}

\begin{figure}
\epsscale{1.1}
\plotone{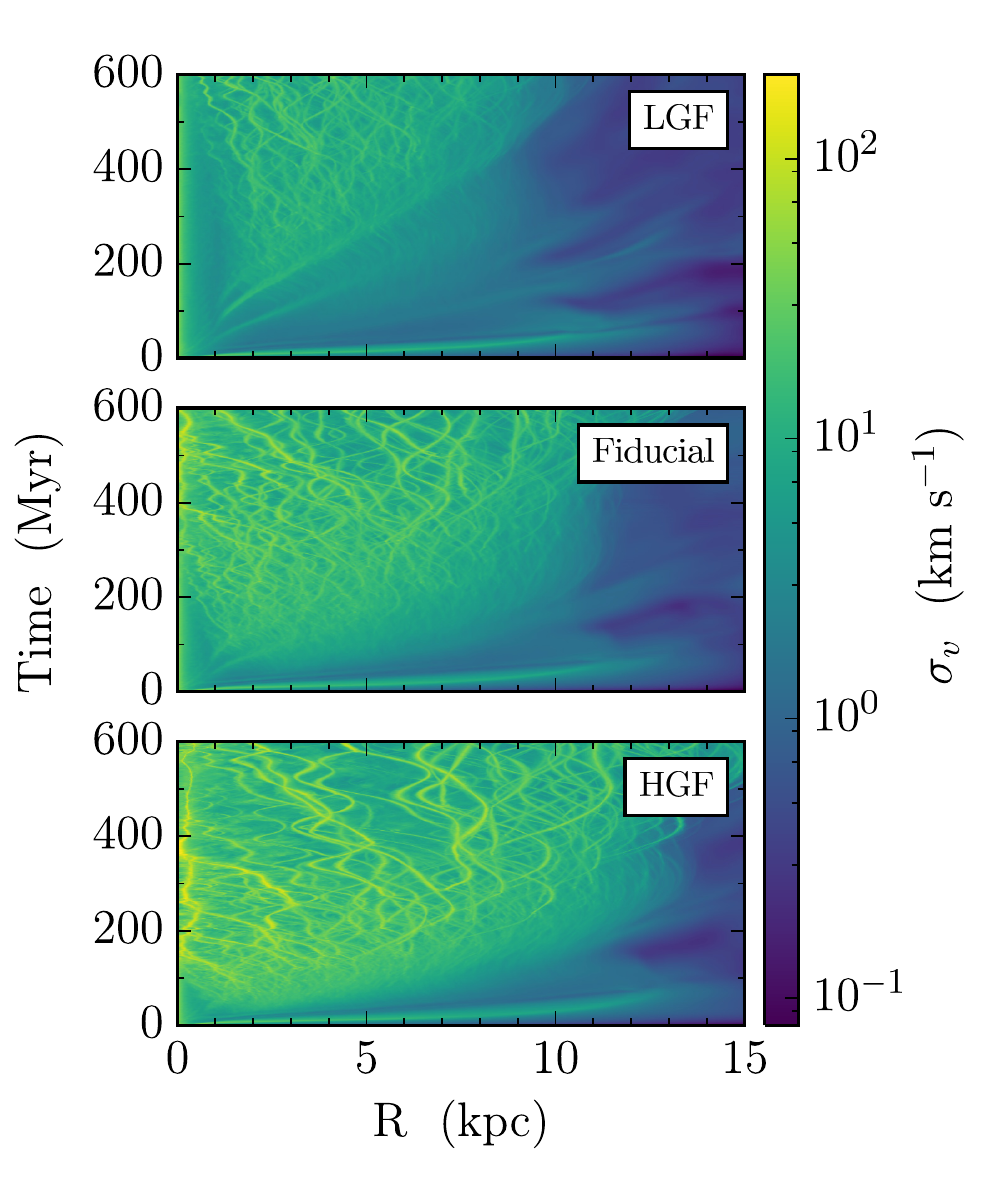}
\caption{The time evolution of the azimuthally averaged velocity dispersion
  for the gas in each of our simulations. Gravitational instability alone is
  sufficient to drive substantial ($\unit[>10]{km\ s^{-1}}$) velocity
  dispersions in all three simulations.}
\label{sigmav_evolution}
\end{figure}

Here we examine the detailed gas velocity structure in our simulated
galaxies. We are particularly interested in the radial dependence in the
effective sound speed, the anisotropy in the velocity dispersion, and the
relative contribution of turbulent motions and thermal sound speed to the
effective sound speed. In \autoref{sigmav_evolution} we present the time
evolution of the azimuthally averaged effective sound speed in each of our
simulations. In all three simulations, gravitational instability alone is able
to drive substantial turbulent velocity dispersions in the inner disk. The
outer, gravitationally stable disk is characterized by much lower velocity
dispersions. We also note that the typical velocity dispersion scales with the
initial gas fraction. This may be interesting for studies of high redshift gas
disks, which are known to exhibit substantial turbulent velocity dispersions.

To obtain a more complete picture of the velocity structure in our simulations,
we look in more detail at the time-averaged velocity structure in the fiducial
simulation. We perform the same sort of time averages described in
\autoref{time_average_section}, focusing on the velocity dispersion, sound
speed, effective sound speed, and velocity anisotropy.

\begin{figure}
  \epsscale{1.2} \plotone{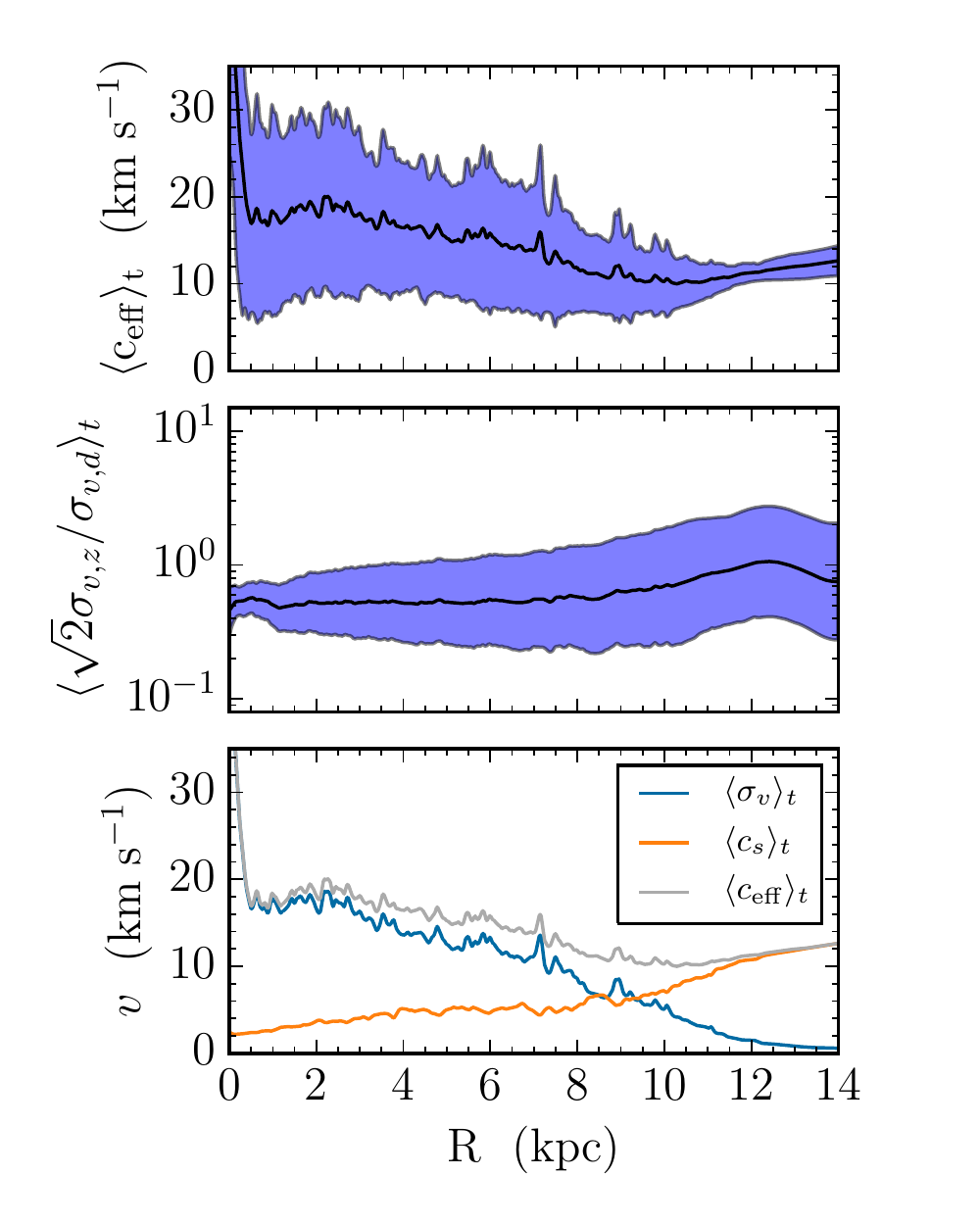}
  \caption{The time-averaged effective sound speed (top and bottom panel) and
    velocity dispersion anisotropy (middle panel) for the gas in our fiducial
    simulation.  Shaded blue regions indicate the 1-$\sigma$ scatter at a fixed
    radius. In the bottom panel, we show the contribution to the effective sound
    speed due to the velocity dispersion and sound speed. In the inner disk,
    bulk velocity dispersion dominates the effective sound speed, while at large
    radii the thermal component dominates. Note that these are mass-weighted
    averages so the hot, low-density interarm medium does not contribute
    significantly to these profiles.\\}
  \label{velocity_summary}
\end{figure}

We present the results of this time averaging in \autoref{velocity_summary}. In
the top panel, we plot the average effective sound speed as a function of
radius, along with the 1-$\sigma$ dispersion in the effective sound speed at any
given radius. The effective speed is typically of order \unit[$\sim\,$20]{km
  s$^{-1}$}, with a gradual radial decline comparable to what is observed in
\hI\ maps of nearby galactic disks (see the discussion and references in
\autoref{motivation}). This is perhaps surprising given that feedback processes
are often invoked as an energy source for turbulence. Instead, we find
generically that gravitational instability alone is sufficient to drive
substantial turbulent velocity dispersions, more than enough to match observed
gas velocity dispersions in resolved observations of galactic gas kinematics.

Since the averages used in these profiles are mass weighted, the hot low-density
interarm regions do not contribute significantly. As can be seen by inspecting
\autoref{evolve_summary}, the interarm gas has sound speeds of order
$\unit[100]{km\ s^{-1}}$, but as we show in \autoref{velocity_summary}, a
typical gas parcel (in a mass-weighted sense) has an effective sound speed of
order $\unit[20]{km\ s^{-1}}$. We make use of mass-weighted averages so that the
measured effective sound speed profiles are representative of the kinetic energy
content of the gaseous disk.

Our galaxy models look quite different from real galaxies
when we look at the anisotropy in the velocity dispersion components. To
investigate this, in the middle panel of \autoref{velocity_summary} we present
the time-averaged, azimuthally averaged ratio of the vertical velocity
dispersion to the in-plane velocity dispersion in the fiducial galaxy model,
$\sqrt{2} \sigma_{v,z} / \sigma_{v,d}$. We include the factor of $\sqrt{2}$ in
the numerator so that a value of unity would correspond to a flow field with
isotropic turbulence, in which the dispersion in the two in-plane components of
the velocity field account for exactly twice as much kinetic energy as the one
out-of-plane component. In our model galaxies, the we find that this ratio is
approximately $0.5$ over the bulk of the disk, indicating that in-plane motions
contribute substantially more to the turbulent kinetic energy. This can also be
seen in the thickness of our disks, where we typically find that the scale
height is of order one or two cell spacings on the maximum refinement level. The
extreme 2D disks we form are likely an artifact of the lack of feedback in our
simulations.

Finally, we show in the bottom panel of \autoref{velocity_summary} how the
effective sound speed we measure in the inner regions of the gaseous disk is
primarily due to bulk turbulent motions. The small sound speed we measure is
typical of the dense gas in our simulation, which cools to temperatures of
\unit[$\sim\,$200--300]{K} due to metal line cooling. The large velocity
dispersions we measure indicates that the dense gas participates in large-scale
turbulent motions.

\section{Discussion and Conclusions}

\label{discussion}

\subsection{Turbulence in a Gravity-Dominated ISM}

Our simulations show, in agreement with previous numerical studies by
\citet{agertz09, agertz15} and \citet{bournaud10}, that gravitational
instability is capable of driving turbulence and stabilizing a disk galaxy at
$Q\gtrsim 1$ even in the absence of any additional energy input from star
formation. This state is characterized by a velocity dispersion that decreases
somewhat with radius from \unit[20]{km s$^{-1}$} near galactic centers to
\unit[10]{km s$^{-1}$} at large radii, in good agreement with observed gas
velocity dispersions in nearby disk galaxies \citep{tamburro09,
  ianjamasimanana15}. Indeed, the only possibly-significant disagreement between
the kinematic behavior found in our simulations and that observed in nature is
that our galaxies' velocity dispersions are anisotropic by a factor of $\sim 2$,
leading to a scale height that is too small. Nonetheless, our results suggest
that, even if stellar feedback is required to explain the vertical velocity
distribution of the ISM, it need not be the dominant energy source for all
turbulent motions. Gravity alone can generate the required in-plane motion.

It is interesting to consider our simulations in light of the arguments commonly
made that star formation feedback is the dominant driver of turbulence in
galaxies. The primary observational argument for feedback driving turbulence is
that there is a correlation between star formation rate and velocity dispersion,
both within galaxies \citep{tamburro09} and from galaxy to galaxy
\citep{green10}.  However, our simulations would also display precisely such a
correlation: the regions of our galaxy that have the highest velocity dispersion
are also the most strongly star-forming, and our models with the highest gas
fractions show both the highest star formation rates and the highest velocity
dispersions.  The point is not that feedback cannot drive turbulence, simply
that a correlation between star formation and turbulence does not necessarily
imply a causal relationship between the two. A correlation of this sort can be
produced even when there is no feedback.

The fact that galaxies can maintain $Q\gtrsim 1$ and continuously drive strong
turbulence without star formation feedback calls into question analytic models
in which galaxies' star formation rates are set by the need to maintain $Q=1$
\citep[e.g.,][]{thompson05, faucher-giguere13}. We find that galaxies can
maintain $Q\gtrsim 1$ regardless of the level of star formation
feedback. Feedback is still needed to produce star formation rates in agreement
with observation and drive out-of-plane motions, but the condition that
determines the level of feedback and star formation appears to be completely
decoupled from the need to maintain $Q\gtrsim 1$.

\subsection{Fueling Star Formation, From $z\sim\, 2$ to Today}

The primary result of our simulations is that gravitational instability-driven
turbulence is capable of inducing significant bulk mass flows in galaxies,
leading to a migration of mass inward from the passive outer regions of disks
toward their actively star-forming centers. The inward mass transport rates we
measure in our simulated galaxies are comparable to the star formation rates of
typical $L_*$ galaxies. While there has been a great deal of work done on such
inward migration in the context of the observed giant clumps in $z\sim\, 2$
galaxies \citep[e.g.,][]{genzel08, cresci09, dekel09b, bournaud09}, our
simulations show that a completely analogous phenomenon can operate for Milky
Way-like galaxies at $z = 0$.

An important implication of galaxies' ability to transport gas inward at rates
comparable to their star formation rates is that their available fuel for star
formation is their full gas reservoir, not simply the material in the actively
star-forming inner disk. This is significant because, while gas depletion times
are much less than the Hubble time only considering inner disk gas, the same is
not true for many galaxies at $z=0$ if we consider the full gas reservoir. For a
volume-limited sample of local star-forming galaxies with stellar masses in the
range $\log (M_*/M_\odot) = 10 - 11.5$, \citet{saintonge11} find typical \hI\
depletion times of $\sim\, 3$ Gyr, with no strong dependence on stellar mass,
and \hmol\ depletion times that range from $\sim\, 0.5 - 3$ Gyr from the lowest
to the highest stellar masses in the sample. If galaxies have access to their
full extended \hI\ reservoirs to fuel star formation, then their total depletion
times are simply the sum of these, implying that galaxies with stellar masses
above $\sim\, 10^{11}$ $M_\odot$ have depletion times of $\sim\,
6$ Gyr. This is still less than the Hubble time, but not by much. Galaxies in
this mass range could receive no new gas supply after $z\sim\, 1$ and still fuel
all their present star formation, particularly once the contribution from
stellar recycling is included. Additionally, although galaxies with masses in 
 the range $\log (M_*/M_\odot) = 10 - 11$ have shorter molecular gas depletion 
times, the atomic gas depletion times are also $\sim \unit[3]{Gyr}$. If these
galaxies are also able to transport atomic gas from their outskirts into the
active star-forming centers, they could supply all of their star formation
without accreting any gas at all since $z \sim 0.5$.

We are therefore forced to conclude that strong equilibrium between gas inflow,
star formation, and outflows may, by $z=0$, exist only for dwarf galaxies with
$M_* \lesssim \unit[10^{9.5}]{M_\odot}$, since dwarfs are able to eject mass out
of their star forming regions via high mass-loading factor winds. This means the
mass range $M_* \sim 10^{9.5} - 10^{10.5}$ $M_\odot$, which corresponds to where
galaxies are most efficient at converting baryons to stars \citep{behroozi13},
may not necessarily be in a state of near-instantaneous equilibrium between gas
accretion, star formation, and outflows. Galaxies with masses above
$M_* \sim 10^{9.5} M_\odot$ may be far from equilibrium, contradicting one of
the central assumptions of ``bathtub'' models of galaxy formation
\citep[e.g.,][]{bouche10, lilly13, forbes14b, mitra15}.

\acknowledgments\

This work utilized the Hyades supercomputer at the University of California
Santa Cruz, which is supported by the National Science Foundation through award
AST-1229745, and the Pleiades supercomputer, which is supported by the NASA
Advanced Supercomputing Division. The computations and analysis described in
this paper rely heavily on open source software packages, including
\texttt{Enzo}, \texttt{Python}, \texttt{yt}, \texttt{IPython}, \texttt{NumPy},
\texttt{SciPy}, \texttt{matplotlib}, \texttt{Cython}, \texttt{hdf5},
\texttt{h5py}, and \texttt{numexpr}.  We thank the developer communities of
these packages for their contributions. This work was supported by NSF graduate
fellowships (NJG and JCF), by NSF grants AST-0955300, AST-1405962 (MRK, NJG, and
JCF), and ACI-1535651 (NJG), by NASA TCAN grant NNX14AB52G (MRK, NJG, and JCF),
by Hubble Archival Research grant HST-AR-13909 (JCF and MRK), and by the Gordon
and Betty Moore Foundation's Data-Driven Discovert Initiative through Grant
GBMF4651 to Matthew Turk.

\begin{appendix}

\section{Analysis}

\label{analysis}

Here we describe the analysis pipeline used to post-process our simulations.

\subsection{Grid Slabs}

\label{grid-slabs}

Rather than analyzing the AMR data structures directly, we instead perform the
bulk of our analysis using data interpolated onto uniform resolution meshes
with \unit[20]{pc} cell spacings encompassing the galactic disk out to a radius
of 20 kpc and to a height of \unit[1.25]{kpc} above and below the disk. This
choice significantly eases the implementation of the various analysis tasks we
want to perform on the data without introducing a significant amount of
error. Since the disk is very thin, a vertical extent of only \unit[1.25]{kpc}
comfortably encloses the gas and stellar disk at all times.

To interpolate the gas data onto a uniform resolution grid, we make use of the
smoothed covering grid object in \texttt{yt}. This operation uses cascading
trilinear interpolation to represent an AMR dataset at a uniform resolution.
The fields representing gas density, the velocity vector components, and the gas
thermal energy were extracted from the raw simulation outputs. Since the
gravitational potential is not normally written to disk, we made use of the
\texttt{-g} command-line option of the \texttt{Enzo} code to solve the Poisson
equation using the gravity in \texttt{Enzo} during post-processing.

So that we can apply the same analysis tasks we use for gas fields defined on
\texttt{Enzo}'s AMR mesh to N-body star particles in our datasets, we opt to
analyze stellar fields by depositing the star particle data onto the uniform
resolution grids slabs. In all cases, we use cloud-in-cell interpolation onto
grids with the same shape and resolution as those used for the gas data. Our
choice of uniform resolution slabs leads to issues with data sparseness in the
outskirts of our simulated galaxies but is a good match for the density of
particle data within a radius of \unit[10]{kpc}.

\subsection{Rotation Curve and Epicyclic Frequency}

Because our simulations use a live stellar and dark matter halo, the rotation
curve of our galaxy is not fixed, and must instead be computed self-consistently
from the simulation outputs. Following \citet{binney08} and \citet{shu92}, we
note that for an axisymmetric system with gravitational potential field $\Phi(R,
z)$, we can define the circular frequency
\begin{equation}
\Omega^2(R) = \frac{1}{R} {\left( \frac{\del \Phi}{\del R} \right)}_{(R, 0)},
\end{equation}
and epicyclic frequency
\begin{equation}
\kappa^2(R) = \frac{2 \Omega}{R} {\left( 2 R \Omega + R^2 \frac{d \Omega}{d
      R}\right)}.
\end{equation}
This implicitly assumes the disk is thin so we can infer the circular velocity
curve by only considering the gravitational potential in the midplane.

To measure the rotation frequency, we extract the gravitational potential in a
slice at the midplane of the galaxy, and evaluate the partial derivative of the
potential with respect to $x$ and $y$ coordinates using a centered finite
difference of the gravitational potential. We then form a 2D array of the
gradient with respect to the cylindrical $R$ coordinate out of the images of the
$x$ and $y$ gradients. This results in a local estimate of the rotation
frequency based on the local radial gradient in the gravitational potential. To
average over local departures from axissymmetry, we create our final estimate of
the rotation frequency by fitting a spline interpolator to a binned version of
our local estimate of the circular frequency as a function of radius. This
produces a binned \unit[20]{pc} resolution estimate of the rotation frequency as
a function only of cylindrical radius. Finally, we calculate the circular
velocity via
\begin{equation}
v_c = R \Omega.
\end{equation}

\subsection{Surface Density}

Several different quantities we are concerned with are defined in terms of the
projection of our simulation data. For a 3D density field $\rho(x, y, z)$ we can
define the surface density,
\begin{equation}
\Sigma(x, y) = \int_{-\infty}^{\infty}\rho \,dz.
\end{equation}
Here $\rho$ represents the mass density of gas or stars, which we denote as
$\rho_{\rm gas}$ and $\rho_*$ below. Both quantities are defined on the uniform
resolution grid slabs discussed in \autoref{grid-slabs}.

Discretizing $\rho$ into a uniform resolution 3D array, which we denote as
$\rho_{i j k}$, the continuous definition of the surface density reduces to
\begin{equation}
\Sigma_{i j} = \sum_{k=0}^{N_z} \rho_{ijk} \Delta z
\end{equation}
where $\Delta z$ is the cell spacing the vertical direction. Since our grid
slabs only include data within \unit[1.25]{kpc} of the disk midplane, we are
implicitly assuming that gas well off the midplane does not contribute
significantly to the surface density.

We introduce the notation, $\langle q \rangle_z$, to represent the discrete
mass-weighted projection operator.  The result is a uniform resolution 2D array,
where the $i, j$ resolution element can be found by computing,
\begin{equation}
  \langle q \rangle_{z, ij} = \frac{1}{\Sigma_{ij}} \sum_{k=0}^{N_z} \rho_{ijk}
  q_{ijk} \Delta z.
\end{equation}
This notation allows us to write several of the definitions below in a compact
form.

\subsection{Velocity Dispersion}

\label{velocity_dispersion}

Since turbulent motions are a significant component of the energy budget in the
ISM, we would like to directly measure the turbulent kinetic energy in our
simulations. Since our simulations are a discretized representation of a
continuous underlying system, we can only estimate the velocity dispersion by
comparing velocity values in cells contained within a moving window several
cells across.  So that variations in the rotation curve over the extent of the
window do not tend to inflate the measured velocity dispersions, we define the
velocity of streaming motions
\begin{equation}
\delta \mathbf{v} = \mathbf{v} - v_c(R)\hat{\phi}.
\end{equation}

To simplify notation in our definition of the velocity dispersion, we define the
discrete operator
\begin{equation}
C_{ijk}(x) =
\sum_{k^\prime=k-2}^{k+2}\sum_{j^\prime=j-2}^{j+2}\sum_{i^\prime=i-2}^{i+2}
x_{i^\prime j^\prime k^\prime},
\end{equation}
where $x$ is a field defined on the uniform resolution mesh. This is equivalent
to a discrete convolution of $x$ using a cubical top-hat kernel function. The
top-hat kernel includes 5 cells, so the spatial scale of the convolution is
\unit[100]{pc}. Using this notation, our estimate of the velocity dispersion due
to motions in the $x$ coordinate, $\delta v_x$, can be written
\begin{equation}
  \sigma_{x, ijk} = \frac{\sqrt{C(\rho)C(\rho \, {\delta v_x}^2) - C{(\rho \,
        \delta v_x)}^2}}{C(\rho)},
\end{equation}
where $\delta v_x$ is the x component of the streaming velocity vector
$\delta \mathbf{v}$.  The expressions for the $y$ and $z$ velocity dispersions
are analogous. We write the variance in terms of the convolution operator since
this form lends itself to faster one-pass parallel reduction. This result can be
derived from the definition of the weighted variance for a sampled quantity with
measurements $x_i$ and weights $w_i$, and weighted mean $\bar{x}_w$,
$\sigma^2 = \sum_{i=1}^N w_i \left( x_i - \bar{x}_w \right)^2 / \sum_{i=1}^{N}
w_i$
by expanding the squared term and substituting for the definition of the
weighted mean.

We use the same definition for stellar velocity dispersions, but rather than
calculating the dispersion on a cell-by-cell basis, we instead iterate over
particles, performing a running dispersion calculation by depositing particle
densities and velocities into accumulating arrays.  This uses more memory, but
is substantially faster for the particle counts in our simulations. Like the gas
velocity dispersions, the stellar velocity dispersion is computed on a spatial
scale of \unit[100]{pc}.

Finally, we calculate the projected velocity dispersions by performing a
weighted projection of the turbulent kinetic energy density:
\begin{equation}
\sigma_v = \sqrt{\langle \sigma_x^2 + \sigma_y^2 +
  \sigma_z^2 \rangle_z}.
\end{equation}
In addition, to estimate the relative contribution of in-plane and out-of-plane
turbulent motions, we separately define the in-plane velocity dispersion
\begin{equation}
\sigma_{v,d} = \sqrt{\langle \sigma_x^2 + \sigma_y^2 \rangle_z},
\end{equation}
and the out-of-plane velocity dispersion
\begin{equation}
\sigma_{v,z} = \sqrt{\langle \sigma_z^2 \rangle_z}.
\end{equation}

\subsection{Effective Sound Speed}
\label{ceff}

We can define an effective sound speed that takes into account both gas pressure
and turbulent pressure,
\begin{equation}
\ceff = \sqrt{\sigma_v^2 + c_s^2}
\end{equation}
For collisionless fluids, $\ceff = \sigma_v$, but for the gaseous component we
must calculate the sound speed. In practice, we do this in terms of the weighted
projection of the thermal energy density $e$,
\begin{equation}
c_s = \sqrt{\gamma(\gamma -1) \langle e \rangle_z}
\end{equation}
where $\gamma =5/3$ is the adiabatic index.

\subsection{Toomre $Q$}
\label{toomre_q}

Using the derived data we introduced above, we can compute the Toomre $Q$
parameter for both the gas,
\begin{equation}
Q_{\rm gas} = \frac{c_{\rm eff}\kappa}{\pi G \Sigma_{\rm gas}}
\end{equation}
and stars,
\begin{equation}
Q_* = \frac{\sigma_{v,*} \kappa}{3.36 G \Sigma_*}.
\end{equation}

We also attempt to estimate the degree of gravitational instability due to both
the gas and stars.  We make use of the formula derived by \citet{romeo11} that
takes into account the separate contribution of the gas and stars as well as
their finite thickness.  In this formalism,
\begin{equation}
\label{romeoq}
\frac{1}{Q_{\rm total}} = \left\{
    \begin{array}{lr}
      \frac{W}{T_*Q_*} + \frac{1}{T_{\rm gas}Q_{\rm gas}} & : T_*Q_* \ge T_{\rm gas}Q_{\rm gas}\\
      \frac{1}{T_*Q_*} + \frac{W}{T_{\rm gas}Q_{\rm gas}} & : T_*Q_* < T_{\rm gas}Q_{\rm gas}
    \end{array}
    \right.
\end{equation}
where $T = 0.8 + 0.7 \sigma_z/\sigma_d$, and $W =
2\sigma_*\sigma_{\rm{gas}}/(\sigma_*^2 + \sigma_{\rm gas}^2)$.

\subsection{Scale Height}

For a fluid with local mass density $\rho$, we define the scale height $h$ such
that
\begin{equation}
\int_{0}^h \rho \, dz = \left(1 - e^{-1}\right) \int_{0}^{\infty} \rho\, dz
\end{equation}
Here the $z$ coordinate is perpendicular to the disk and centered on the disk
midplane. Since the fluid distribution is not necessarily symmetric in $z$, the
$z$ coordinate of the midplane is not constant, and the scale height
above and below the midplane might not be equal. In practice, we
take the true scale height $h$ to be the arithmetic mean of the scale heights
measured above and below the midplane.

We generate maps of of the scale height by individually processing $z$-aligned
pencil stacks of cells. We calculate a running sum of the gas mass along each
pencil stack and note the $z$ locations where the surface density equals one of
${(2e)}^{-1} \Sigma$, $\Sigma/2$, and $(1.0 - {(2e)}^{-1})\Sigma$.  We use
linear interpolation in $z$ to estimate the intra-cell locations where the
running sum exceeds each critical value. Finally, we calculate a single scale
height estimate by averaging the ``top'' and ``bottom'' estimates.

\subsection{Radial Mass Flux}

For a cylindrical test volume $V$ of radius $R$, the flux of mass across the
surface of the cylinder is
\begin{equation}
\dot{M} = R \int_0^{2\pi} \int_{-\infty}^\infty  \rho v_{r} \, dz \, d\theta.
\end{equation}
In practice, we calculate this quantity using a discrete approximation based on
our interpolated data defined on uniform resolution grid slabs. Here we can take
advantage of the cylindrical symmetry of the problem to substantially reduce the
computational cost of this calculation. If we consider a single $x$-$y$ slice
through our grid slab, it is a straightforward geometric problem to find the set
of cells in this slice that the cylindrical test volume intersects.

If we define a one dimensional index $l \in [0, N]$ where $N$ is the total
number of cells in the $x$-$y$ plane that intersect with the circular test
region, we can approximate
\begin{equation}
\dot{M} \approx  \sum_{k=0}^{k_{\rm max}}\sum_{l=0}^{N}\rho_{lk} v_{r, lk}
\Delta\theta_{l}\Delta z.
\end{equation}
Here $l$ maps to a set of unique coordinates in the 2D slice of the 3D grid slab
and $k$ indexes along the $z$ direction in the grid slab. One can think of $l$
as a map to a single pair of $i, j$ indices for the $x$ and $y$ rows in our grid
slice. The problem of calculating the radial mass flux reduces to calculating
the angles subtended by each cell $\Delta \theta_l$. Since we ignore gas well
off the midplane, this implicitly assumes that the mass transport at large
heights above and below the midplane is small.

Doing this at many different radii allows up to map how the radial mass flux
varies with galactocentric radius. In practice, we choose a set of 1000 nested
test cylinders, binned evenly in radius with a spacing of \unit[40]{pc}.

\subsection{Time Averaging}

\label{time_average_section}

We are interested in the time-average behavior of azimuthally averaged
quantities, both to compare to 1D models of galaxy formation and also to capture
the large-scale behavior of our simulations over long periods of time. As we
show in the main text, the internal structure of our simulated galaxies is
initially dominated by transient disturbances. After several galactic rotation
periods, the disks settle down and reach a quasi-equilibrium state in which the
global structure is more or less static. Since the galactic rotation period is
an increasing function of radius, the places where we expect the disk to be
settled corresponds to a wedge shaped region of radius-time phase space.

For any azimuthally-averaged quantity $x(r,t)$, we define its time average, a
function of radius only, as
\begin{equation}
  \langle x(r,t) \rangle_t = \frac{1}{t_{\rm max} - (t_0 +
    ar)}\int_{t_0+ar}^{t_{\rm{max}}}x(r,t) dt
\end{equation}
where $t_{\rm max}$ is the simulation time at the end
of the simulation (\unit[600]{Myr} for all three simulations presented here) and
$a$ and $t_0$ are constants that define the shape of the ``averaging wedge'' as
denoted by the blue lines in \autoref{radius_time_mass_flux}. In
practice we determine these constants ``by eye'' based on inspection of the
simulation results.

\end{appendix}

\bibliography{ms}

\end{document}